\pdfoutput=1 
\documentclass[%
 reprint,
superscriptaddress,
 amsmath,amssymb,
 aps,
prb,
floatfix,
nofootinbib
]{revtex4-1}

\usepackage[final]{changes}
\definechangesauthor[name={Albert}, color=orange]{per}
\usepackage{lipsum}
\usepackage[utf8]{inputenc}
\usepackage{graphicx}
\usepackage{float}
\usepackage{blkarray}
\usepackage{amsmath}
\usepackage{url}
\usepackage{caption}
\usepackage{subcaption}
\usepackage{xcolor}
\usepackage{multirow}
\usepackage{xspace}
\usepackage{xfrac}
\usepackage{acro}

\usepackage[OT2,T1]{fontenc}
\usepackage[inline]{enumitem}

\DeclareFontFamilySubstitution{OT2}{\rmdefault}{wncyr}

\DeclareAcronym{FCM}{short=FCM, long=Force Constant Matrix}
\DeclareAcronym{MLIP}{short=MLIP, long=Machine Learning Interatomic Potential}
\DeclareAcronym{MLMD}{short=MLMD, long=Machine Learning accelerated Molecular Dynamics}
\DeclareAcronym{NDSC}{short=NDSC, long=Non-Diagonal Supercell}
\DeclareAcronym{DSC}{short=DSC, long=Diagonal Supercell}
\DeclareAcronym{BZ}{short=BZ, long=Brillouin Zone}
\DeclareAcronym{LAMMPS}{short=\texttt{LAMMPS}, long=Large-scale Atomic/Molecular Massively Parallel Simulator \cite{LAMMPS}}
\DeclareAcronym{GAP}{short=GAP, long=Gaussian Approximation Potential}
\DeclareAcronym{ACE}{short=ACE, long=Atomic Cluster Expansion}
\DeclareAcronym{SOAP}{short=SOAP, long=Smooth Overlap of Atomic Positions}
\DeclareAcronym{DFT}{short=DFT, long=Density Functional Theory}
\DeclareAcronym{SCF}{short=SCF, long=Self-Consistent Field}
\DeclareAcronym{RMSE}{short=RMSE, long=Root Mean Squared Error}
\DeclareAcronym{MAE}{short=MAE, long=Mean Average Error}
\DeclareAcronym{PES}{short=PES, long=Potential Energy Surface}
\DeclareAcronym{MD}{short=MD, long=Molecular Dynamics}
\DeclareAcronym{EC}{short=EC, long=Einstein Crystal}
\DeclareAcronym{NETI}{short=NETI, long=Non-Equilibrium Thermodynamic Integration}
\DeclareAcronym{IP}{short=IP, long=Interface Pinning}
\DeclareAcronym{GGA}{short=GGA, long=Generalized Gradient Approximation}
\DeclareAcronym{hcp}{short=hcp, long=hexagonal close packed}
\DeclareAcronym{bcc}{short=bcc, long=body-centred cubic}

\newcommand{\abinitio}{\textit{ab initio\xspace}}
\newcommand{\etal}{\textit{et al\xspace}}

\begin{document}
\title{Multi-Phase Dataset for Ti and Ti-6Al-4V}
\author{Connor S. Allen}
\affiliation{Department of Physics, University of Warwick, Coventry CV4 7AL, United Kingdom}
\author{Albert P. Bart\'ok}
\affiliation{Department of Physics, University of Warwick, Coventry CV4 7AL, United Kingdom}
\affiliation{Warwick Centre for Predictive Modelling, School of Engineering, University of Warwick, Coventry CV4 7AL, United Kingdom}

\begin{abstract}
\begin{center}
Titanium and its alloys are technologically important materials that display a rich phase behaviour.
In order to enable large-scale, realistic modelling of Ti and its alloys on the atomistic scale, \acfp{MLIP} are crucial, but rely on databases of atomic configurations.
We report databases of such configurations that represent the $\alpha$, $\beta$, $\omega$ and liquid phases of Ti and the Ti-6Al-4V alloy, where we provide total energy, force and stress values evaluated by \ac{DFT} using the PBE exchange-correlation functional.
We have also leveraged and extended a data reduction strategy, via non-diagonal supercells, for the vibrational properties of Ti and sampling of atomic species within bulk crystalline data for Ti-6Al-4V.
These configurations may be used to fit \ac{MLIP} models that can accurately model the phase behaviour of Ti and Ti-6Al-4V across a broad range of thermodynamic conditions.
To validate models, we assembled a set of benchmark protocols, which can be used to rapidly develop and evaluate \ac{MLIP} models.
We demonstrated the utility of our databases and validation tools by fitting models based on the \ac{GAP} and \ac{ACE} frameworks.\\

\vspace{0.2 cm}
\small{
UK Ministry of Defence © British Crown Owned Copyright 2024/AWE-NST
}
\end{center}
\end{abstract}

\maketitle

\section{Introduction}
The transition metal Ti and its ternary alloy Ti-6Al-4V (Ti 90 wt\%, Al 6 wt\%, V 4 wt\%) have industrial relevance in aerospace, biomedical, defence and high-performance engineering applications due to the machinability, anti-corrosive and high strength-to-weight properties of the material \cite{gurrappa_characterization_2003,veiga_properties_2012}.
Pure Ti has been experimentally observed to form in the $\alpha$ (\ac{hcp}, P$6_3$/mmc) phase at ambient conditions, undergoing a transformation to the $\beta$ (\ac{bcc}, Im-3m) at approximately 1150 K \cite{zhang_experimental_2008, dewaele_high_2015, young_phase_2023} at ambient pressure.
First-principles modelling shows the $\omega$-Ti (hexagonal, P6/mmm) phase to be the ground state \cite{joshi_stability_2002,kutepov_crystal_2003,ahuja_titanium_2004,verma_high-pressure_2007,hao_first-principles_2008,mei_density-functional_2009,hao_first-principles_2010,dewaele_high_2015}.
The relative stability of Ti at high pressure has also been investigated extensively by both computational \cite{joshi_stability_2002,kutepov_crystal_2003,ahuja_titanium_2004,verma_high-pressure_2007,hao_first-principles_2008,mei_density-functional_2009,hao_first-principles_2010,dewaele_high_2015} and experimental \cite{xia_crystal_1990,tonkov_high_1992,vohra_novel_2001,akahama_new_nodate,ahuja_titanium_2004,errandonea_pressure-induced_2005,ponosov_optical_2012,dewaele_high_2015} studies, with significant controversy surrounding the phase transition boundary of $\alpha \rightarrow \omega$ and the existence of high pressure phases.
Experimental studies have established that Ti-6Al-4V forms a polycrystallineThus far, DFT structure at ambient conditions, predominantly of the $\alpha$ structure with the existence of interspersed $\beta$ grains between phase boundaries, of which contain large concentrations of V and reduced Al content \cite{tan_revealing_2016}.
Similarly to pure Ti, the associated phase diagrams of Ti-6Al-4V reveal that the predominant solid phases of interest of this alloy include the $\alpha$, $\beta$ and $\omega$ phases \cite{macleod_phase_2021,kalita_ti-6al-4v_2023}. 

\acp{MLIP} have recently emerged as surrogate models of the \abinitio{} Born-Oppenheimer \ac{PES} that retain first-principles accuracy at a very moderate computational cost and linear scaling with system size \cite{Schuett.2018,Shapeev.2016, drautz_atomic_2019,Behler.2017,Unke.2019,bartok_gaussian_2010}. 
These models are built by carrying out non-linear regression to reproduce microscopic observables obtained from \abinitio{} calculations, e.g. total energies, forces and stresses, as a function of the atomic positions.
The mapping of atomic positions is usually encoded such that a given atomic environment is invariant under translations, rotations, and permutations of identical species, where these encoders are often called descriptors, symmetry functions or fingerprints.
Examples include the \ac{SOAP}\cite{bartok_representing_2013} and \ac{ACE}\cite{drautz_atomic_2019}.
These surrogate models can maintain \abinitio{} accuracy whilst significantly reducing the computational cost associated with evaluating an atomic configuration, when compared against the underlying \abinitio{} method used to develop the \ac{MLIP}. 

Our main result is that we constructed databases of atomic configurations, labelled by \abinitio{} calculations, of Ti and Ti-6Al-4V representing multiple thermodynamically stable phases below 30 GPa for use in \ac{MLIP} development.
We have also developed a set of benchmarks that may be used to validate any \ac{MLIP} fitted using our database.
Finally, we utilised the \ac{GAP} and \ac{ACE} frameworks to fit \acp{MLIP} to demonstrate the coverage and sufficiency of our database.

\section{Methodology}
\subsection{Density Functional Theory}\label{sec:Ti64_DFT}
The \abinitio{} calculations that provided the labels (total energies, forces and stresses) in the training database and reference benchmarks were performed using the plane-wave \ac{DFT} code, \verb|CASTEP|\ (v24.1)\cite{castep}.
On-the-fly ultrasoft pseudopotentials were also generated for Al, V and Ti with respective valence electronic structures: 3s$^2$ 3p$^1$, 3s$^2$3p$^6$3d$^2$4s$^2$, and 3s$^2$3p$^6$3d$^2$3s$^2$.
The PBE  \cite{perdew_generalized_1996} functional was used to approximate exchange-correlation.
Parameters of \ac{DFT} calculations are set such that our calculations are converged to sub-meV/atom relative to a computationally excessive basis.
We have found that this level of convergence can be achieved by applying a plane-wave energy cut off of 800 eV, and sampling the electronic \ac{BZ} using a Monkhorst-Pack grid that had a spacing of 0.02 \AA$^{-1}$.

To obtain reference atomic configurations, geometry optimisations were performed for the experimentally observed symmetries of Ti below 12 GPa with a maximum force tolerance of less than 1 meV/\AA\, a stress tolerance of less than 0.1 GPa, and energy tolerance of $10^{-9}$ eV/atom for \ac{SCF} cycles.
This provided the relaxed lattice parameters for each crystalline phase, which was later used for generating training data and benchmark calculations.
The geometry relaxations of the physically relevant phases in pure Ti were also used as the basis for constructing the Ti-6Al-4V dataset. \\

\subsubsection{Training database}
Representing the exact stoichiometry of the Ti-6Al-4V alloy would require the minor alloying components represent 10.2\% Al and 3.6 \% V by number. To approximate this stoichiometry within 2\%, one may construct approximate primitive cells for each crystalline phase being considered.
For the $\alpha$ and $\beta$ phases the smallest such configuration, containing at least 1 V atom per unit cell, contains 28 atoms ($3\times2\times2$ and $4\times3\times2$ respectively) with 3 Al and 1 V (10.7\% Al and 3.6 \% V by number), and for the $\omega$ phase the a 24 atom supercell ($2\times2\times2$) may be considered with 3 Al and 1 V (12.5\% Al and 4.2 \% V).
The maximise of the coverage of the database, we considered a random substitutions of Ti by the other alloying elements. 
In order to save computational resources, we aimed to construct a dataset using a minimal number of atoms per \ac{DFT} calculation in a given periodic cell, whilst sampling the possible disorder, both vibrational and substitutional, efficiently.
This then motivated the \ac{NDSC} approach \cite{allen_optimal_2022,lloyd-williams_lattice_2015} to constructing crystalline configurations for \abinitio{}  database building, and this extended version of the \ac{NDSC} strategy is outlined in \ref{sec:ti64_database}.\\

\subsubsection{Validation database}
For validation, we created larger periodic unit cells of atomic configurations such that we can represent a more realistic disorder of the minor alloying components in a simulation cell.
The size of these atomic structure models, for each crystalline phase, were chosen to be as large as tractably possible for calculation of \ac{DFT} labels.
These configurations are intentionally left out of the training data and we refer to them as the validation dataset when evaluating energies, forces and virial stresses of the developed \acp{MLIP} against our reference \ac{DFT}. 

We use these benchmark configurations to evaluate the \ac{NDSC} strategy, which requires relatively small unit cells to constructing data points for medium-entropy crystalline systems.
The performance of \ac{MLIP} models fitted using the \ac{NDSC} data can be validated against the \ac{DFT} predictions using these atomic configurations. 

Utilising the orthorhombic ground state geometries of pure Ti as a starting point, we first generated supercells for the $\beta$ ($4\times4\times4$, 128 atoms), $\alpha$ ($3\times3\times3$, 108 atoms) and $\omega$ ($2\times3\times3$, 108 atoms) crystalline structures.
From these atomic configurations, we also generate isotropic volume perturbations of 96\% and 98\% of the ground state volume per atom of pure Ti for crystalline system, providing data points to study transferability of \acp{MLIP} to high pressure.
The atomic positions within the unit cells are then displaced from ideal lattice sites according to a normal distribution with standard deviation of $0.10$ \AA, such that each volume perturbation had 3 such samples.
The atomic species were set randomly such that we recover the stoichiometry of Ti-6Al-4V in each configuration to within 1 \%. 
In our benchmarks, we evaluate the surrogate \ac{MLIP} on these configurations and compare against the \ac{DFT} labels.

\subsubsection{Elastic behaviour}
In order to evaluate the response of the \acp{MLIP} to cell deformation, we compute the elastic constants for each crystalline symmetry with \ac{DFT} as a benchmark.
In these calculations we construct supercells containing $3\times3\times3$ ($\alpha$, $\omega$) and $4\times4\times4$ ($\beta$) repeating units of the primitive cell for each crystalline phase.
In these configurations we substitute the alloying components using the special quasirandom structures \cite{zunger_special_1990} algorithm within the integrated cluster expansion toolkit \cite{angqvist_icet_2019} \verb|python| library, on the approximate unit cells.
The cell vectors and atomic positions of these structures were relaxed using \ac{DFT}, and elastic constants were fit utilising the finite differences method implemented within the \verb|matscipy| package\cite{grigorev_matscipy_2024}. \\

\subsubsection{Vibrational properties}
To calculate our reference phonon dispersions and density of states, firstly geometry optimisations were performed on each atomic structure until a structure with maximum force of less than 1 mev/\AA\ is found, with the stress change tolerance below 0.1 GPa, with an energy tolerance of $10^{-9}$ eV used for \ac{SCF} cycles.
We calculated the force constant matrices utilising the finite displacement method within \verb|CASTEP|, using a finite displacement of $0.02$ \AA{} ($0.01$ \AA{} for pure Ti), in supercells corresponding to a uniform $4\times4\times4$ grid in the vibrational \ac{BZ}.
We calculate the phonon dispersions along high symmetry lines for each crystalline symmetry \cite{Setyawan.2010}.
Phonon density of states was calculated on a uniform $40\times 40\times 40$ $\mathbf{q}$ grid of the vibrational \ac{BZ}.

Calculating phonon dispersions requires the replication of a simulation cell so that one can accurately sample between high symmetry points in the vibrational \ac{BZ}.
In the case of pure Ti, this is easily tractable with \ac{DFT}, as the primitive cell structures for each crystalline phase contain only a few atoms, however, utilising the approximate primitive simulation cells of Ti-6Al-4V would have required excessive computational effort, as the unit cells need to be larger to accommodate the stoichiometry of the alloy.
For this reason, when evaluating phonon dispersions relevant to Ti-6Al-4V, we instead calculate the phonon dispersions, density of states, and harmonic free energies (for dynamically stable structures) for a series of smaller supercells (no more than 12 atoms) in which for every crystal symmetry a Al-Al, Al-V and V-V interaction as nearest neighbours is being considered.
Inevitably, these configurations contain a significantly higher ratios of the alloying elements, nevertheless, they provide valuable benchmark data against which \ac{MLIP} models can be compared.

Interactions between Al and V atoms occupying nearby atomic sites have been found to be an energetically favourable \cite{dumontet_elastic_2019}, and our previous results indicate that Al-Al ordering is likely disfavoured.
We utilise these benchmark configurations specifically to assess the performance of surrogate \ac{MLIP} models on capturing the interactions of minor alloying components, as similar configurations are more sparsely represented within in the training data.\\


\subsection{Database Generation}\label{sec:ti64_database}
\subsubsection{Multiphase Ti}
As plane wave DFT scales excessively with the number of atoms simulated ($\mathcal{O}(N^3)$), to capture each crystalline system effectively we utilised \acp{NDSC}\cite{lloyd-williams_lattice_2015} as a basis for representing the vibrational properties and substitutional disorder.
This strategy allows for a much more efficient sampling of the vibrational \ac{BZ}, and as a result, requires less computational effort to achieve a given accuracy \cite{allen_optimal_2022} for a fitted \acp{MLIP} model.
We considered \acp{NDSC} commensurate with the sampling achieved using $4\times 4 \times 4$ supercells of the primitive unit cells.
This results in no more than 4 repeating units of the primitive unit cell for each crystal symmetry, efficiently limiting the total number of atoms required in the \ac{DFT} calculations.
For each \ac{NDSC}, volume perturbations were generated by isotropically straining the cell such that points along the pressure axis are uniformly sampled.
The configurations then had the atomic coordinates randomly perturbed by a normal distribution with standard deviation of $0.10$ \AA.
Additional data was generated for the $\alpha$ and $\omega$ phases that utilise the same volume perturbations, however, normally distributing atomic positions around ideal lattice sites with a standard deviation of  $0.02$ \AA.
We also capture anisotropic deformations of the unit cell.
This was achieved by generating symmetric strain tensors, $\boldsymbol{\epsilon}$, which is used to transform the lattice cell vectors as $\mathbf{L}_{\textrm{rand.}} = (\mathbf{I} + \boldsymbol{\epsilon})\mathbf{L}_0$, where $\mathbf{L}_0$ are the original cell vectors of the simulation cell.
The entries of this strain matrix are generated from the uniform distribution $\epsilon_{i\leq j}\sim \mathcal{U}(-0.01, 0.01)$, and internal atomic coordinates were also scaled with the cell deformation such that atoms remained at the same fractional coordinate. 

To augment our database, we added disordered atomic configurations representing the liquid state.
While our intention was not necessarily a thermodynamically accurate sampling of the configurational space of the liquid, we used molecular dynamics to ensure that the collected sample configurations are thermodynamically relevant.
To efficiently sample the liquid phase of Ti, we utilise the \acf{MLMD} \cite{stenczel_machine-learned_2023} feature of the \verb|CASTEP| package.
In \ac{MLMD}, both \ac{DFT} and \acp{MLIP} are used to calculate the \ac{PES} at given points in the molecular dynamics trajectory depending on a \ac{PES} calculator selection algorithm.
\verb|CASTEP| uses the \ac{GAP} framework to achieve significant acceleration of \abinitio{} molecular dynamics calculations without compromising the accuracy of the trajectories.
With this framework, \ac{GAP} surrogate models are generated on the fly in an automatic fashion.
In particular, a \ac{MLMD} simulation may be started with the first few time steps being integrated using forces obtained from \ac{DFT}, while storing these labels in a database to train a \ac{GAP} model.
The algorithm switches between computationally expensive \ac{DFT} and cheap \ac{MLIP} evaluations adaptively, using \ac{DFT} labels to retrain the surrogate model when necessary.,
Alternatively, one may also provide a training database prior to starting a \ac{MLMD} simulation, where to this database is then appended further \ac{DFT} evaluations in the \ac{MLMD} trajectory.

The \ac{MLMD} approach accelerates \abinitio{} molecular dynamics as the number of time steps in a given simulation can be significantly increased, allowing the sampling of more configurations in the relevant thermodynamic phase space.
We considered supercells 54 and 128 atoms for $\beta$-Ti, and generated volume perturbations by isotropic scaling the lattice parameter between 102 \% and 90\% of the ground state value.
To initialise the atomic positions into a disordered state, we perform molecular dynamics using the \ac{LAMMPS} package with the Ti1 EAM potential by Mendelev \etal \cite{mendelev_development_2016} in the NVT ensemble.
We first overheated the crystalline structure via setting the thermostat to 4000~K, followed by quenching to approximately 2000~K, and retaining atomic configurations to serve as initial geometries for \ac{MLMD}.
Each \ac{MLMD} trajectory begins by first performing 5 initial \abinitio{} steps, then a \ac{GAP} model is trained and 10 further steps are computed with the surrogate model.
After this, the accuracy of the surrogate model is checked against a full \abinitio{} calculation, and is re-trained with the incorporation of the new \abinitio{} data.
The switching algorithm we utilise in \ac{MLMD} is such that we adaptively change the number of steps between error checks.
When the surrogate model passes the success criterion, the check interval is doubled, while the interval is halved in the case of unsuccessfully fulfilling the accuracy criterion. We consider a minimum of 10, and a maximum of 100, surrogate steps between checks.
We evaluate the success of the surrogate as having less than 5~meV/atom error relative to the \abinitio{} calculation.
The model complexity utilised in the \ac{GAP} surrogate consisted of a 2-body kernel with 20 sparse points and cutoff distance of 4.5~\AA, and a many-body \ac{SOAP} descriptor with 1200 sparse points with the following descriptor hyperparameters for many body interactions: $n_{\textrm{max}}=8$, $l_{\textrm{max}}=8$, $r_{\textrm{cutoff}}=6.0$ \AA, $\zeta=4$, and $\sigma_{\textrm{atom}}=0.5$\AA.
In \ac{MLMD}, we utilise the Nose-Hoover thermostat with time constant of 200 fs in the NVT ensemble with a timestep of 2 fs. Across the 54 atom configurations in the \ac{MLMD} trajectories a mean of 2.06~ps of simulation time was considered over 9 independent trajectories.
For the 128 atom configurations, the mean simulation time was  1.81~ps over 4 independent trajectories.

\begin{figure*}[h!tb]
    \centering
    \includegraphics[width=0.35\linewidth]{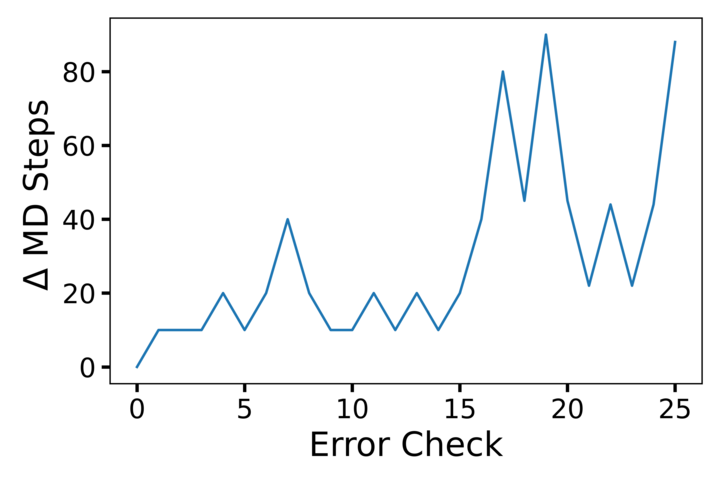}
    \includegraphics[width=0.35\linewidth]{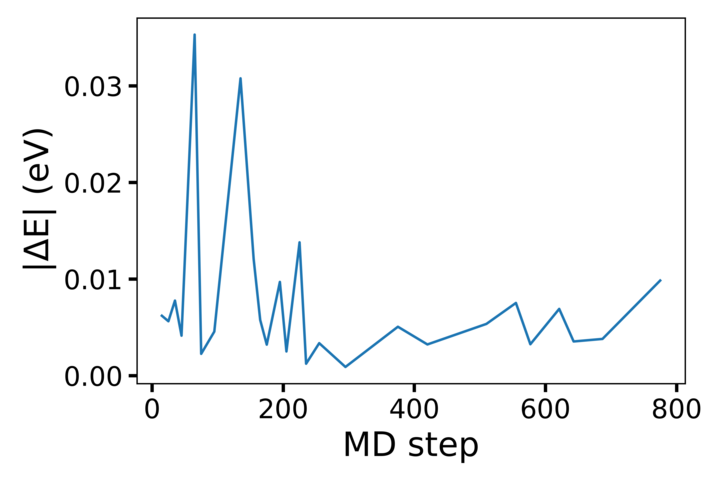}\\
    \includegraphics[width=0.35\linewidth]{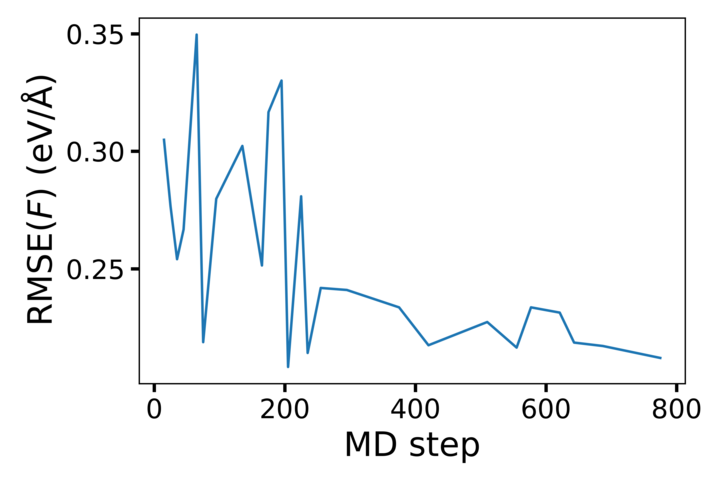}
    \includegraphics[width=0.35\linewidth]{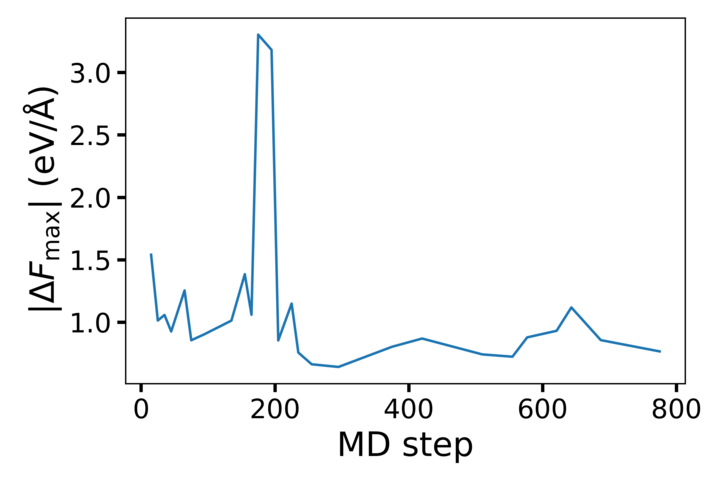}\\
    \includegraphics[width=0.35\linewidth]{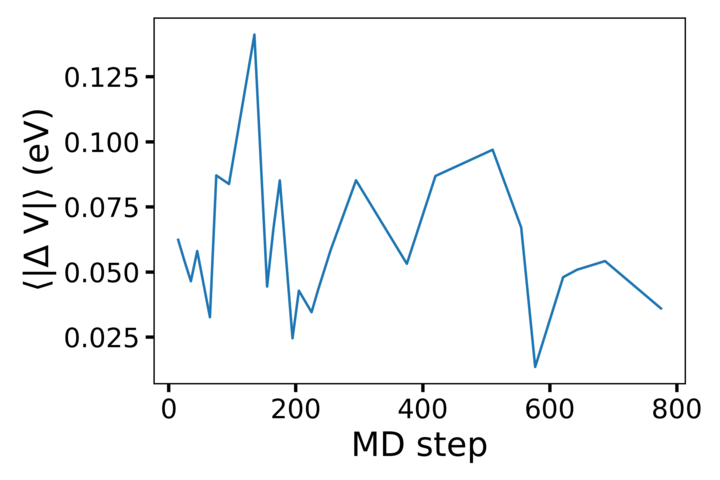}
    \includegraphics[width=0.35\linewidth]{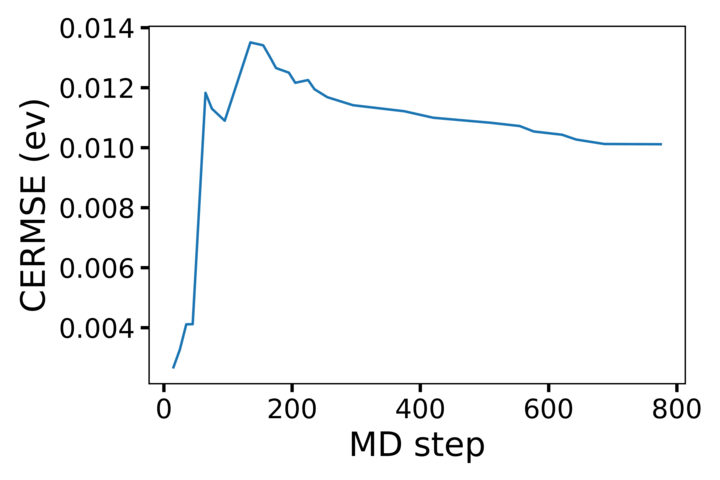}\\
    \includegraphics[width=0.35\linewidth]{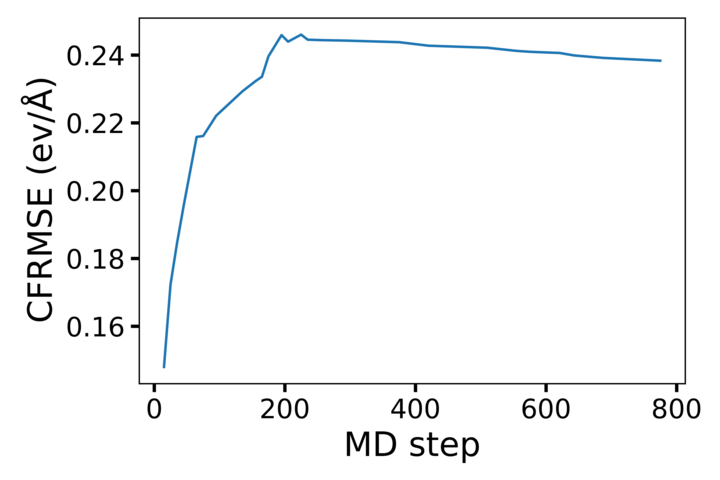}
    \includegraphics[width=0.35\linewidth]{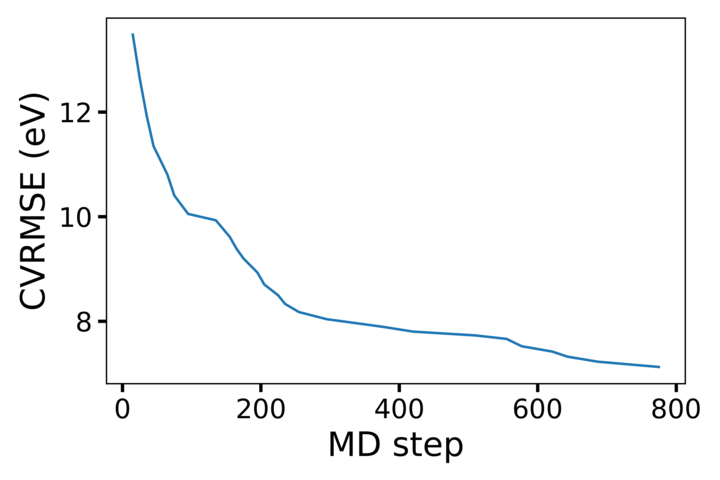}
    \caption{\raggedright Learning rates of various quantities of interest of surrogate model for liquidus Ti-6Al-4V of during accelerated \abinitio{} molecular dynamics. From left to right, top to bottom, we present the following properties: number of molecular dynamics (MD) steps between \abinitio{} and surrogate calculator checks, difference between configuration energy per atom, \ac{RMSE} of atomic forces, absolute error of maximum error discrepancy, average absolute difference in virial stress, cumulative \acs{RMSE} of the configuring energy per atom (CERMSE),  cumulative \acs{RMSE} of the atomic forces (CFRMSE), and  cumulative \acs{RMSE} of the virial stress (CVRMSE).}
    \label{fig:apx_ti64_liquid_learning_curves}
\end{figure*}

\subsubsection{Ti-6Al-4V}\label{sec:generation_Ti64}
For Ti-6Al-4V, we utilise and extend the \ac{NDSC} method for generating bulk crystalline data \cite{allen_optimal_2022, lloyd-williams_lattice_2015}, by considering chemical perturbations on the cell similarly to vibrational \ac{BZ} sampling.
In this scheme we generate a series of \acp{NDSC} for each crystalline symmetry with varying levels of vibrational \ac{BZ} sampling.
Initially isotropic volume perturbations were considered by generating a set of \acp{NDSC} with a grid sampling of $8\times8\times8$ for each physically observed symmetry of pure Ti.
This resulted in 44, 56, and 44 starting configurations for $\alpha$, $\beta$ and $\omega$ respectively.
From these, configurations containing less than 4 ($\alpha$ and $\beta$) and 6 ($\omega$) atoms were removed as the data being constructed was concerned with targeting a dilute regime of minor alloying elements.
Isotropic volume perturbations were generated by scaling the lattice parameters in the range of 95\% to 102\% of the ground state pure Ti geometry in 1\% increments.
Atomic positions were then perturbed around ideal lattice sites via a normal distribution with standard deviation of $0.10$~\AA{} for each volume perturbed \ac{NDSC} for a total of 6 samples.
The atomic species of the isotropic volume perturbed \acp{NDSC} were then randomly swapped from Ti to Al and/or V. For configurations in the $\alpha$ phase, we consider up to 1, 2 and 3 atomic species swaps from Ti to Al and/or V for configurations containing 4, 8 and 12 atoms respectively. For the $\beta$ phase we consider up to 1 and 2 atomic species swaps for configurations of 4 and 8 atoms respectively, and, up to 1, 2 and 4 for $\omega$ phase configurations of 6, 12 and 24 atoms respectively. The number of atomic swaps considered is selected as a random integer on the bounds of 1 to the maximum number of swaps allowed for that configuration type.
As the stoichiometry of Ti-6Al-4V has a larger proportion of Al to V in the alloy, the random sampling of chemical perturbations to minor alloying components in our workflow was biased for a 3:1 (Al:V) ratio.
In total we generated  2064 ($\alpha$), 2496 ($\beta$), and 2064 ($\omega$) configurations for the isotropic volume perturbed dataset, of which 6491 were successfully evaluated with \ac{DFT}.



To capture the elastic properties of Ti-6Al-4V, configurations under shear deformations were generated using \acp{NDSC} as a templates.
In this instance, we utilised the following vibrational \ac{BZ} sampling for each crystalline symmetry: $6\times6\times6$ $(\alpha)$, $12\times12\times12$ $(\beta)$,$4\times4\times4$ $(\omega)$.
From these, we filter the set such that we retain only the 12 atom \ac{NDSC} configurations for each crystalline system, thus resulting in 19 ($\alpha$), 48 ($\beta$) and 8 ($\omega$) starting geometries.
Configurations with atomic positions perturbed from ideal lattice sites are then generated for each set with 16 ($\alpha$), 6 ($\beta$) and 38 ($\omega$) realisations for each \ac{NDSC}.
These configurations are then deformed via a symmetric strain tensor, scaling atomic positions, where the samples of each entry are from the uniform distribution $\epsilon_{i\leq j}\sim \mathcal{U}(-0.01, 0.01)$.
Atomic positions were then displaced according to a normal distribution with standard deviation of 0.05 \AA.
Similarly, we also generate randomly deformed and chemically modified \acp{NDSC} for isotropically scaled volume perturbations.
This was done by scaling the lattice parameters randomly on the interval [0.90, 1.02], for a series of copies of each \ac{NDSC} with perturbed atomic positions, from which was then deformed via the strain tensor described previously.
In total, targeted cell deformation data consisted of 1629 configurations, bringing the total crystalline data via this framework to 8120 configurations with 105436 atomic environments.

To gather information on the liquid phase of Ti-6Al-4V, we constructed two pure Ti supercells, $3\times3\times3$ and $4\times4\times4$, in the orthorhombic $\beta$-Ti crystalline symmetry.
Utilising a multiphase pure Ti \ac{GAP} developed previously, molecular dynamics in \ac{LAMMPS} was preformed in the NPT ensemble on both supercells using a 2 fs timestep.
The velocities of atoms were initialised from a normal distribution corresponding to a temperature of 3000 K, this was then quenched via the Nose-Hoover thermostat from 4000 K to 2500 K with a time constant of 1.35 ps over 20 ps, from which the simulation proceeded at 2500 K for an addition 20 ps to equilibrate the system.
After equilibration, a series of liquidus configurations were generated by taking samples in 4 ps intervals for a total of 5 configurations for each supercell at a given pressure.
We considered pressures up to 20 GPa, in 5 GPa intervals, in pure Ti when generating these configurations via the Parrinello-Rahman barostat with time constant 1.75 ps.
From the 54 atom supercell samples we initialise Ti-6Al-4V configurations by randomly replacing Ti atoms with 6 Al and 2 V atomso or 13 Al and 5 V atoms in the case of the 128 atom supercell. 

From these initial liquidus configurations we perform \ac{MLMD} \cite{stenczel_machine-learned_2023} in CASTEP in the NVT ensemble.
We utilise an adaptive approach to switching between \abinitio\ and \ac{MLIP} calculator during a single molecular dynamics simulation with identical switching criterion as we did for the liquid pure Ti, however, with the surrogate model consisting of many-body \ac{SOAP} descriptors with 400 sparse points, per atomic interaction type, with the following descriptor hyperparameters: $n_{\textrm{max}}=6$, $l_{\textrm{max}}=8$, $r_{\textrm{cutoff}}=6.0$ \AA, $\zeta=4$, and $\sigma_{\textrm{atom}}=0.5$\AA.
Across the 54 atom configurations in the \ac{MLMD} trajectories a mean of 1.09 ps of simulation time was considered over 5 independent trajectories. For the 128 atom configurations, the mean simulation time was 0.84 ps over 16 independent trajectories.

\begin{figure}[h!tb]
    \centering
    \includegraphics[width=\linewidth]{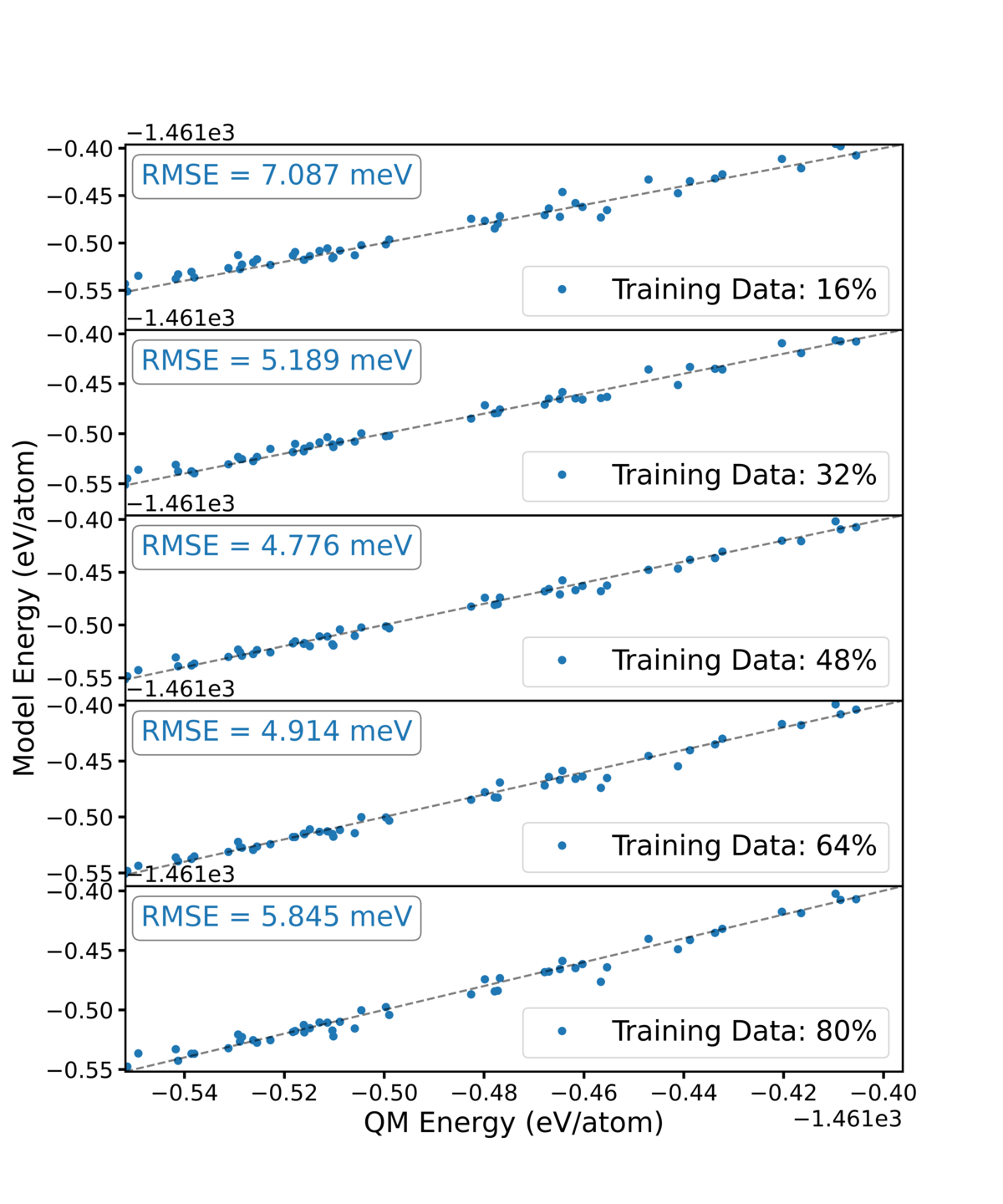}
    \caption[Data saturation for increasing liquidus data.]{\raggedright Model performance on configurational energy for Ti-6Al-4V \ac{GAP} models developed with increasing liquidus data against a validation set.}
    \label{fig:MLMD_data_saturate}
\end{figure}

In order to assess whether enough liquid configurations were collected, we analysed the \ac{MLMD} trajectories and fitted a series of \ac{GAP} models.
Based on our analysis, we expect that regardless of the \ac{MLIP} framework used, the configurations represent an efficient sample of the liquid phase.
Firstly, as an example to demonstrate the efficiency of the \ac{MLMD} approach we present a series of learning curves for different quantities of interest in Figure \ref{fig:apx_ti64_liquid_learning_curves}.
In this \ac{MLMD} trajectory we were simulating 128 atoms of Ti-6Al-4V at 2500 K starting from a configuration corresponding to 0 GPa in pure Ti.
We observe that most of the gains in the accuracy of the surrogate model are achieved from configurations gathered in first quarter of a given \ac{MLMD} trajectory over 1.6 ps.

During the data collection process, we also evaluated the rate of learning of a surrogate across multiple trajectories.
We firstly concatenated all liquidus data, from which we partitioned 20\% as an evaluation set and 80\% as a training dataset.
From the 80\% partition, we train a series \ac{GAP} models for increasing quantity of data as illustrated in Figure \ref{fig:MLMD_data_saturate} and evaluate the \ac{RMSE} on training observables.
We note from Figure~\ref{fig:MLMD_data_saturate} that the improvement with increasing amount of data is non-monotonic such that models with greater than 48\% of available liquidus data no longer provided additional benefit, at least using the same \ac{GAP} model hyperparameters.
The total size of the liquidus Ti-6Al-4V dataset consisted of 304 configurations with 38912 atomic environments.\\

\section{Results}
In order to assess the utility of the generated database, we have fitted a series of \ac{MLIP} models, using the \ac{GAP} and \ac{ACE} frameworks, and evaluated these models on our benchmark data set.
Naturally, the database is not limited to either of these frameworks, and indeed, models with other schemes or more careful hyperparameter tuning may easily outperform the benchmarks presented here.
Our intention is to demonstrate the coverage of the training database and showcase our benchmarks which are intended to be a stringent test of any surrogate model, covering a broad range of thermodynamically relevant conditions.\\

\subsection{Multiphase Ti}

\subsubsection{Fitting Machine Learned interatomic Potentials}
The \ac{GAP} model developed on the multiphase Ti dataset is constructed using a 2-body inverse polynomial kernel alongside the \verb|SOAPTurbo| kernel.
The \ac{GAP} model consisted of a total 3320 sparse points, 3300 of which are contained in the \verb|SOAPTurbo| kernel.
The representative atomic environments for the \verb|SOAPTurbo| kernel were selected from CUR decompositions\cite{Mahoney.2009} of each configuration type within the data set, ensuring a broad coverage of points across each configuration type.
In the case of the 2-body kernel, points were selected uniformly by taking a representative environment from each bin of a histogram of evaluations with the 2-body kernel, described as \verb|uniform| in Klawohn \etal \cite{klawohn_gaussian_2023}.
The 2-body kernel is short range, acting within 3.5 \AA\ with an inverse polynomial basis with exponents -4, -8, -12 and -14, primarily ensuring that the \ac{GAP} model is repulsive at short atomic distances to stop non-physical atomic overlaps such that MD simulations remain stable.
The \verb|SOAPTurbo| kernel serves to capture the many body interactions within the dataset.
We utilise the \verb|SOAPTurbo| kernel hyperparameters\cite{caro_optimizing_2019}): basis complexity of $n_{\textrm{max}}$=8 and $l_{\textrm{max}}$=8 with cut off distances of $r_{\textrm{soft}}$=5.5 \AA, $r_{\textrm{hard}}$=6.0 \AA, atomic-centred Gaussian widths $\sigma_{\parallel} = \sigma_{\perp}$=0.5 \AA, and kernel exponent $\zeta=6$.
The reader is referred to Klawohn \etal \cite{klawohn_gaussian_2023} for further details on the functionality and parameters within of \ac{GAP} and descriptors as implemented in \verb|QUIP|. 

Alongside the \ac{GAP}, an \acf{ACE} \cite{drautz_atomic_2019} is also presented as a computationally lightweight alternative model, implemented via the \verb|ACEpotentials.jl|\cite{witt_acepotentialsjl_2023} suite. Within \verb|ACEpotentials.jl|, the maximum complexity is collapsed to a single number called the \textit{total degree} which can be set independently for each correlation order $\nu$. 
The last step in building an \ac{ACE} potential is using a regression framework to fit coefficients which map input coordinates to observed quantities, in our case \ac{DFT} labels.
Of the multiple choices for regression frameworks implemented within \verb|ACEpotentials.jl|, and the Bayesian linear regression optimiser was utilised in our work.
A series of \ac{ACE} potentials were fit the reproduce the \ac{DFT} labels for combinations of $\nu$ and \textit{total degree}, with $\nu$ ranging from 3 to 5, and \textit{total degree} from 16 to 20.
We observed that utilising  $\nu \geq 4$ resulted in significant over-fitting of the multiphase Ti dataset, and increasing the \textit{total degree} monotonically reduced the training \acp{RMSE} within $\nu=3$.
A spatial cut-off of 6 \AA\ was utilised for all \ac{ACE} models considered. In the final version used $\nu=3$, for a \textit{total degree} of 20 across each correlation order, resulting in a total of 1809 basis functions.

\begin{figure}[h!tb]
     \centering
     \begin{subfigure}[b]{\linewidth}
         \centering
         \includegraphics[width=\textwidth]{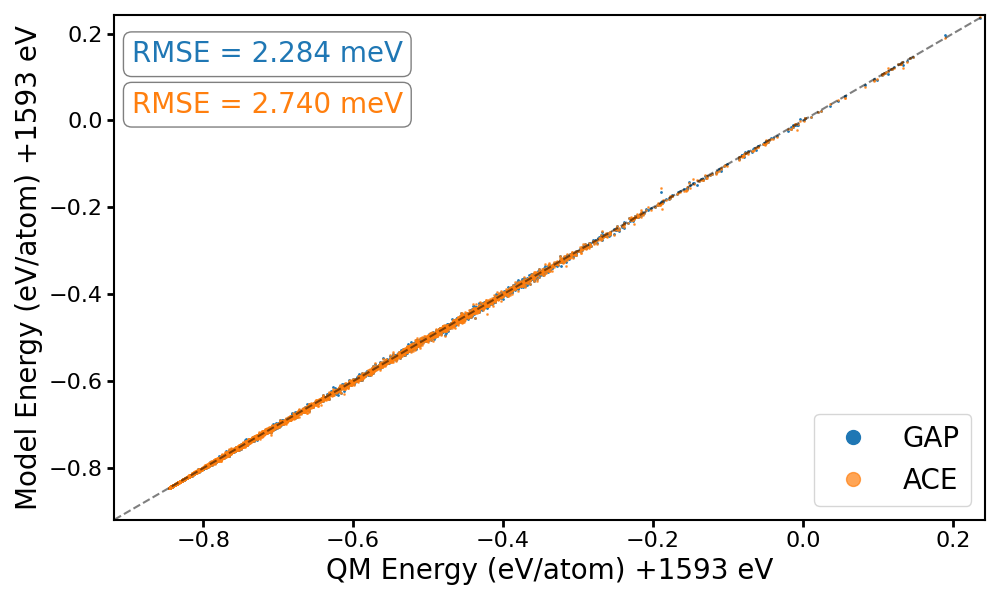}
     \end{subfigure}
     \begin{subfigure}[b]{\linewidth}
         \centering
         \includegraphics[width=\textwidth]{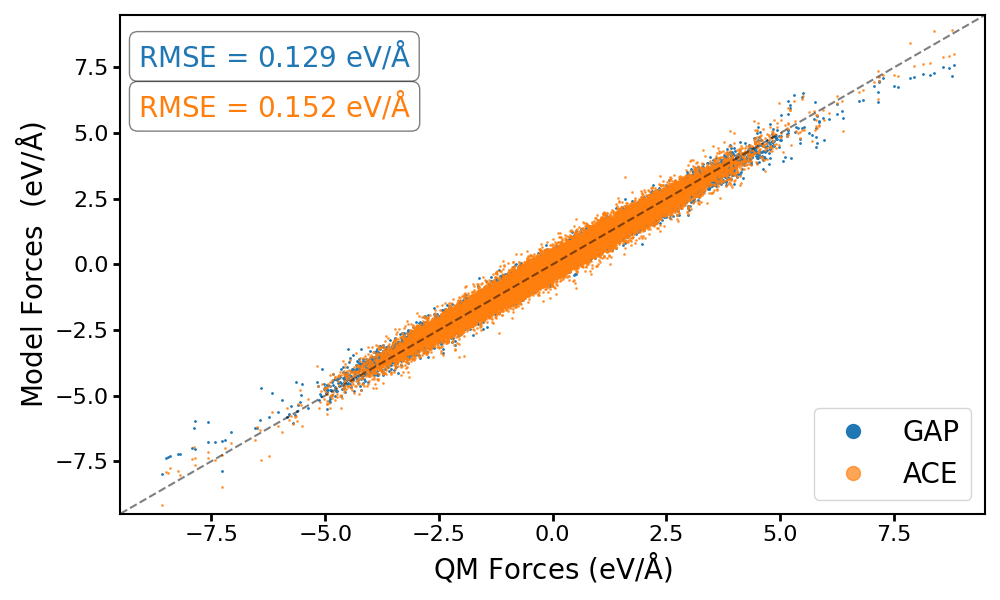}
     \end{subfigure}
     \begin{subfigure}[b]{\linewidth}
         \centering
         \includegraphics[width=\textwidth]{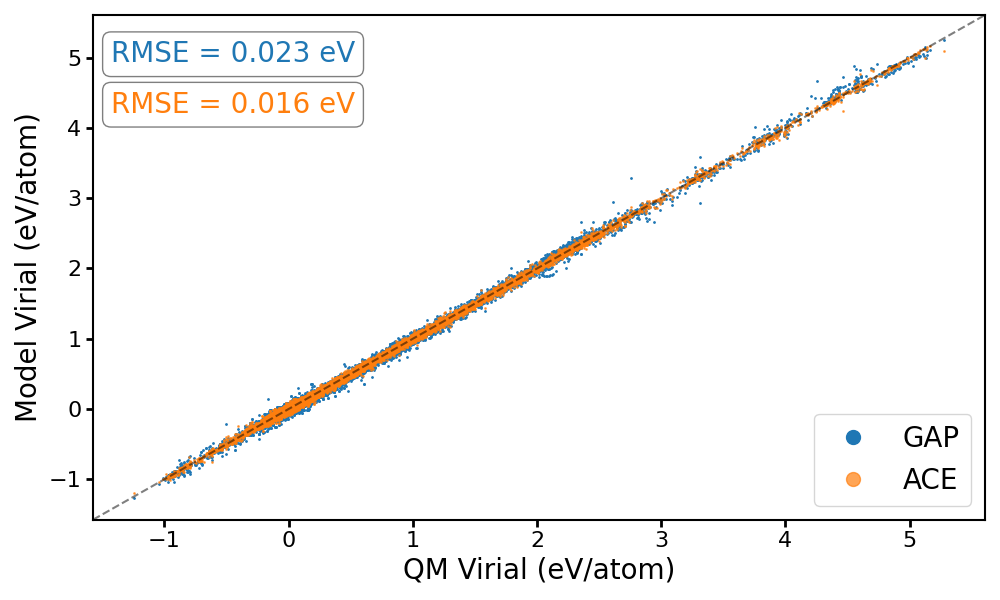}
     \end{subfigure}
        \caption{\raggedright Performance of surrogate models against training observables of the training dataset for each model trained on the multiphase Ti dataset.}
        \label{fig:mTi_RMSE_train}
\end{figure}

After generating the \abinitio{} database through the strategies outlined above, a series of \ac{GAP} and \ac{ACE} models were trained until the final set of hyperparameters for each model was realised. The surrogate models presented here are trained on the total configurational energy, atomic forces and virial stresses, and we evaluate the model performance on reproducing these quantities in Figure \ref{fig:mTi_RMSE_train}.  From the figure, we compare each surrogate model's prediction with that of the \abinitio\ calculation and compute the \ac{RMSE} for each quantity using the entire database.
This constituted 6640 DFT observations, for a total of 82096 atomic environments.
As indicated by Figure \ref{fig:mTi_RMSE_train}, both the \ac{GAP} and \ac{ACE} potential accurately reproduce the underlying data with uncertainty being on the order of meV per atom.\\

\begin{figure*}[h!tb]
    \centering
    \includegraphics[angle=-0,origin=c,width=1.0\linewidth]{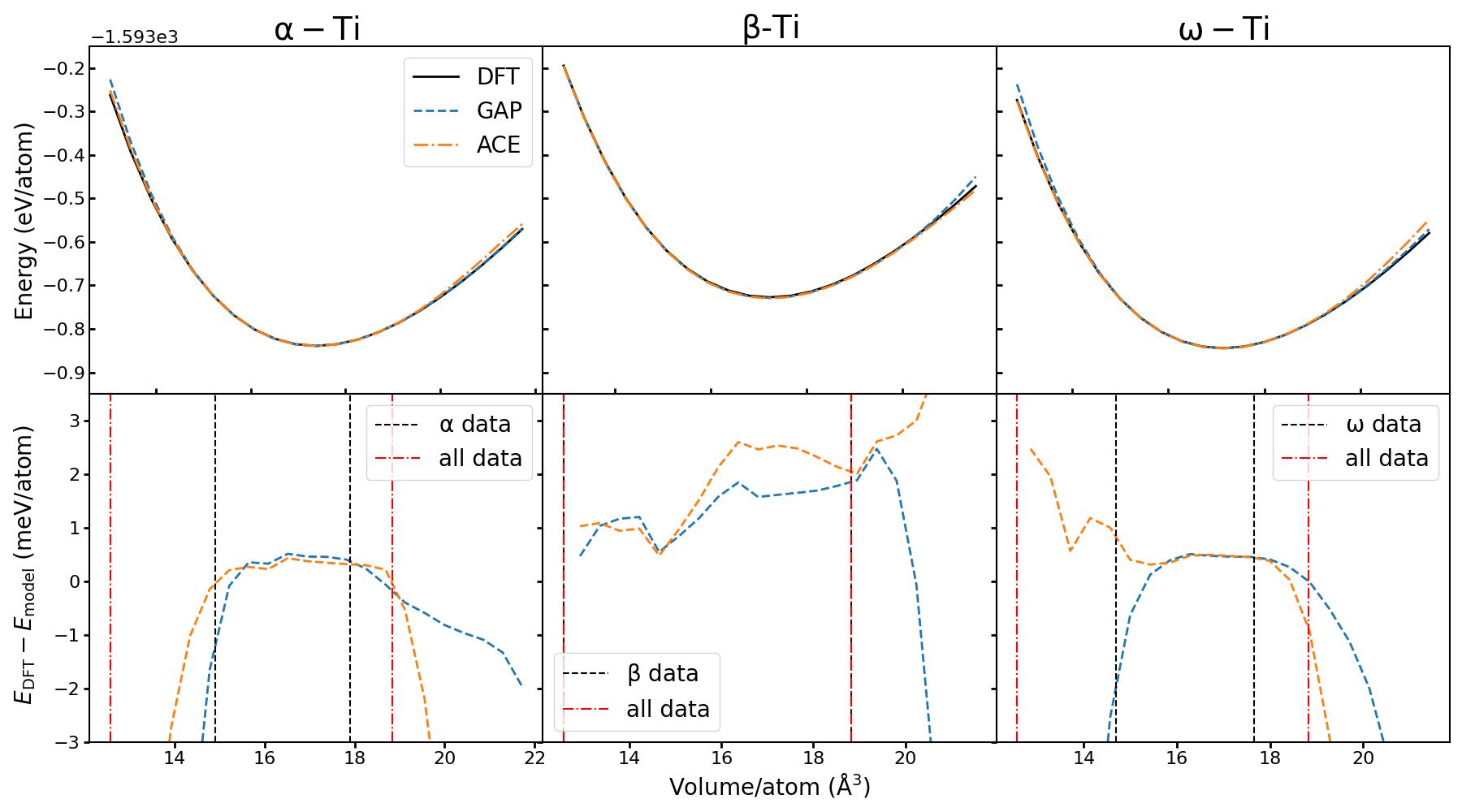}
    \caption[Energy-volume curves in multiphase Ti.]{\raggedright Energy-volume curves (top) and the difference between the surrogate models and underlying DFT (bottom) for each crystal symmetry of Ti. Dashed red lines indicate the bounds where there exists associated crystalline data in the model training, and black dashed indicate that specific to the crystal symmetry.}
    \label{fig:mTi_EV_curves}
\end{figure*}

\subsubsection{Elastic Constants}
We have calculated the lattice parameters and elastic constants of the crystalline phases of Ti using our \ac{GAP} and \ac{ACE} models and compare them to \ac{DFT} benchmark values, presented  in Tables \ref{tab:alphaTi}, \ref{tab:omegaTi} and \ref{tab:betaTi} for $\alpha$-Ti, $\omega$-Ti and $\beta$-Ti respectively.
We find that both surrogate models can reproduce the ground state ambient pressure geometry associated with each crystalline symmetry considered.
As is in agreement with other theoretical investigations in literature \cite{kutepov_crystal_2003,hao_first-principles_2008, mei_density-functional_2009, hu_theoretical_2010, nitol_machine_2022} the \abinitio{} calculations performed in this work find that $\omega$-Ti is the ground state geometry at ambient pressure.
Computed using \ac{DFT}, the energy differences of the phase transitions $\alpha\rightarrow\omega$ and $\alpha\rightarrow\beta$ are $\Delta E_{\alpha\rightarrow\omega} = -5.4$~meV/atom and $\Delta E_{\alpha\rightarrow\beta} = 111.9$~meV/atom, respectively.
Both the \ac{GAP} and \ac{ACE} potentials fitted by us reproduce the energy difference between these crystalline phases as: $\Delta E_{\alpha\rightarrow\omega}^{\textrm{GAP}} = -5.4$ meV/atom and $\Delta E_{\alpha\rightarrow\beta}^{\textrm{GAP}} = -110.6$ meV/atom, and $\Delta E_{\alpha\rightarrow\omega}^{\textrm{ACE}} = -5.5$ meV/atom and $\Delta E_{\alpha\rightarrow\beta}^{\textrm{ACE}} = 109.6$ meV/atom.
In addition, we find that the ground state lattice parameters predicted for each system is in good agreement with previous \abinitio studies at the same level of theory by Mei \etal \cite{mei_density-functional_2009}, Hu \etal \cite{hu_theoretical_2010} and Nitol \etal \cite{nitol_machine_2022}.
The surrogate models also reproduce the elastic properties across the different crystalline symmetries, with particularly good agreement for $\omega$-Ti with a \ac{MAE} across the all elastic constants of 1.5 GPa (1.7\%) and 1.0 GPa (0.9\%) for \ac{GAP} and \ac{ACE} respectively.
The elastic properties of the \ac{DFT} calculations performed here are also in good agreement with value reported in previous literature\cite{hu_theoretical_2010,nitol_machine_2022}.

We also characterise the variation of potential energy variation due to volume perturbations, presented in Figure \ref{fig:mTi_EV_curves}, showing that both the \ac{GAP} and \ac{ACE} potentials accurately capture the bulk modulus of each crystalline system.
To demonstrate how the accuracy of the surrogate models depend on the coverage of the training data, we plotted the difference between the predicted energy compared to the \ac{DFT} data, also indicating the range the atomic volumes present in the training set of configurations.

These energy-volume curves demonstrate that the surrogate models are accurate where there exists the respective training data, as denoted by the hashed boundaries within Figure~\ref{fig:mTi_EV_curves} for a given crystalline symmetry within the crystalline subset of the whole database.
We also observe that for the crystalline phases and immediately around the ground state volume, the energy predicted by both surrogates is somewhat less than that of the reference \ac{DFT} calculation.
In these tests, if an atomic configuration has a specific volume outside of the range represented by the training data, the \ac{GAP} model always predicts the potential energy to be greater than the reference \ac{DFT} in all instances of extrapolation, whereas, the \ac{ACE} model shows in low-volume $\omega$-Ti and high-volume $\beta$-Ti a lower potential energy than the reference.  \\

\begin{table}[h!tb]
    \centering
    \begin{tabular}{||l|c|c|c||}
    \hline
    $\alpha$-Ti &  \hspace{0.2cm} DFT \hspace{0.2cm} &  \hspace{0.2cm} GAP \hspace{0.2cm} &  \hspace{0.2cm} ACE \hspace{0.2cm} \\
    \hline
    Elastic constants (GPa): &&&\\
    \hspace{0.2cm} \footnotesize{$C_{11}$} &180.2 & 173.7 & 175.5 \\
    \hspace{0.2cm} \footnotesize{$C_{33}$} &189.1 & 185.0 & 193.3 \\
    \hspace{0.2cm} \footnotesize{$C_{12}$} &79.0  & 83.5  & 86.7  \\
    \hspace{0.2cm} \footnotesize{$C_{13}$} &76.3  & 79.4  & 73.9  \\
    \hspace{0.2cm} \footnotesize{$C_{44}$} &44.6  & 37.4  & 43.8  \\
    \hspace{0.2cm} \footnotesize{$C_{66}$} &50.6  & 45.1  & 44.4  \\
    \hspace{0.2cm} \footnotesize{$B$}   &112.5 & 113.0 & 112.6 \\
    Lattice parameters: &&&\\
    \hspace{0.2cm} \footnotesize{$a$} (\AA)& 2.939 & 2.938 & 2.938\\
    \hspace{0.2cm} \footnotesize{$c$} (\AA)& 4.647 & 4.648 & 4.648\\
    \hspace{0.2cm} \footnotesize{$V_0$} (\AA$^3$/atom)& 17.38 & 17.38 & 17.38\\
    \hline
    \end{tabular}
    \caption[Elastic constants: $\alpha$-Ti.]{\raggedright Elastic constants and lattice cell parameters for $\alpha$-Ti.}
    \label{tab:alphaTi}
\end{table}

\begin{table}[h!tb]
    \centering
    \begin{tabular}{||l|c|c|c||}
    \hline
    $\omega$-Ti &  \hspace{0.2cm} DFT \hspace{0.2cm} &  \hspace{0.2cm} GAP \hspace{0.2cm} &  \hspace{0.2cm} ACE \hspace{0.2cm} \\
    \hline
    Elastic constants (GPa): &&&\\
    \hspace{0.2cm} \footnotesize{$C_{11}$} &195.0 & 195.0 & 194.6 \\
    \hspace{0.2cm} \footnotesize{$C_{33}$} &247.5 & 244.3 & 243.6 \\
    \hspace{0.2cm} \footnotesize{$C_{12}$} &81.1  & 84.5  & 80.3 \\
    \hspace{0.2cm} \footnotesize{$C_{13}$} &52.7  & 51.6  & 53.8  \\
    \hspace{0.2cm} \footnotesize{$C_{44}$} &54.4  & 53.3  & 54.9  \\
    \hspace{0.2cm} \footnotesize{$C_{66}$} &57.0  & 55.3  & 57.1  \\
    \hspace{0.2cm} \footnotesize{$B$}   &112.0 & 112.1 & 111.8 \\
    Lattice parameters : &&&\\
    \hspace{0.2cm} \footnotesize{$a$} (\AA)& 4.579 & 4.579 & 4.579\\
    \hspace{0.2cm} \footnotesize{$c$} (\AA)& 2.831 & 2.830 & 2.830\\
    \hspace{0.2cm} \footnotesize{$V_0$} (\AA$^3$/atom)& 17.13 & 17.13 & 17.13\\
    \hline
    \end{tabular}
    \caption[Elastic constants: $\omega$-Ti.]{\raggedright Elastic constants and lattice cell parameters for $\omega$-Ti.}
    \label{tab:omegaTi}
\end{table}

\begin{table}[h!tb]
    \centering
    \begin{tabular}{||l|c|c|c||}
    \hline
    $\beta$-Ti &  \hspace{0.2cm} DFT \hspace{0.2cm} &  \hspace{0.2cm} GAP \hspace{0.2cm} &  \hspace{0.2cm} ACE \hspace{0.2cm} \\
    \hline
    Elastic constants (GPa): &&&\\
    \hspace{0.2cm} \footnotesize{$C_{11}$} &91.0 & 87.9   & 86.5   \\
    \hspace{0.2cm} \footnotesize{$C_{12}$} &112.4  & 114.5& 117.0  \\
    \hspace{0.2cm} \footnotesize{$C_{44}$} &40.3  & 45.5  & 39.6   \\
    \hspace{0.2cm} \footnotesize{$B$}   &105.2 & 105.6 &  106.8 \\
    Lattice parameters: &&&\\
    \hspace{0.2cm} \footnotesize{$a$} (\AA)& 3.254 & 3.254 & 3.254\\
    \hspace{0.2cm} \footnotesize{$V_0$} (\AA$^3$/atom)& 17.23 & 17.23 & 17.23\\
    \hline
    \end{tabular}
    \caption[Elastic constants: $\beta$-Ti.]{\raggedright Elastic constants and lattice cell parameters for $\beta$-Ti.}
    \label{tab:betaTi}
\end{table}

\subsubsection{Vibrational Properties}
Another common benchmark within \ac{MLIP} development is how well the \ac{FCM} is reproduced, as this assesses the curvature of the \ac{PES} with respect to atomic displacements.
How well a model reproduces the \ac{FCM} can be represented by presenting the phonon dispersions and the corresponding density of states for each crystalline symmetry of interest, providing quantitative insight into how well a model predicts forces in the harmonic regime of the \ac{PES}.
In our study, the \acp{FCM} were calculated for the developed \acp{MLIP} using the finite difference method \cite{kunc_ab_1982} as implemented in the \verb_phonopy_ package\cite{togo_first_2015}.
We consider the dispersion and density of states for the minima in the \ac{PES} of bulk Ti as represented by $\alpha$-Ti and $\omega$-Ti, and also at the saddle point represented by the dynamically unstable $\beta$-Ti phase.

\begin{figure}[h!tb]
    \centering
    \includegraphics[width=0.8\linewidth]{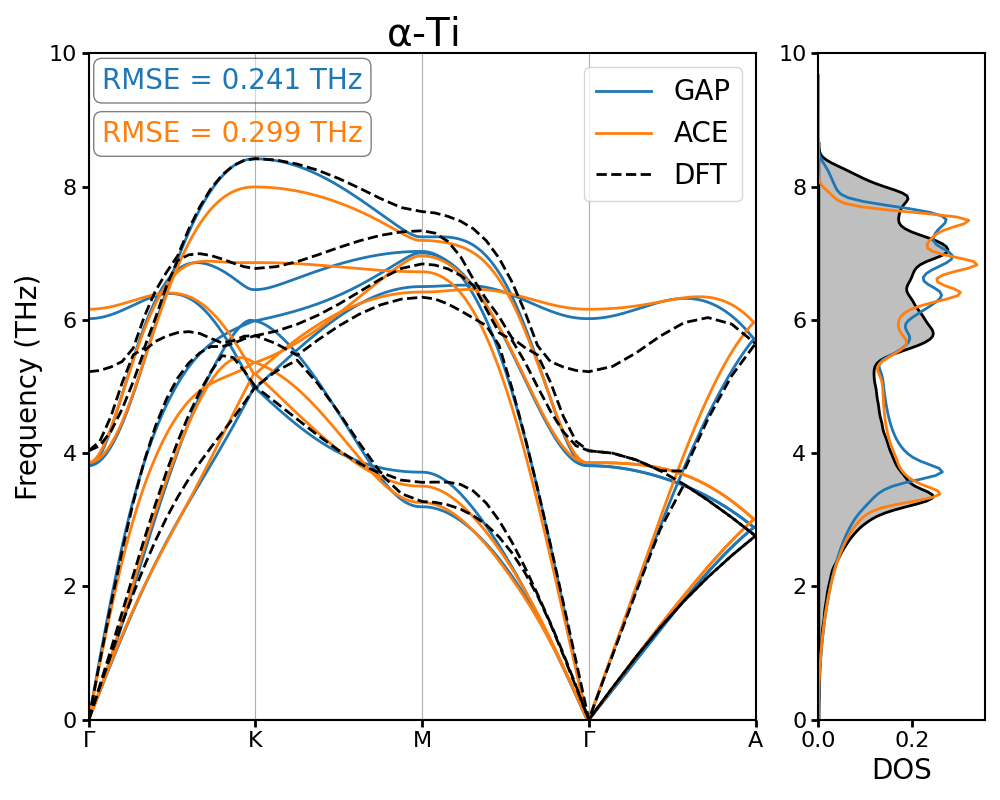}
    \caption[Phonons: $\alpha$-Ti.]{\raggedright Phonon dispersion and density of states for $\alpha$-Ti as calculated by the reference \ac{DFT} calculation alongside \ac{GAP} and \ac{ACE} surrogate models developed.}
    \label{fig:mTi_phonons_hcp_Ti}
\end{figure}

\begin{figure}[h!tb]
    \centering
    \includegraphics[width=0.8\linewidth]{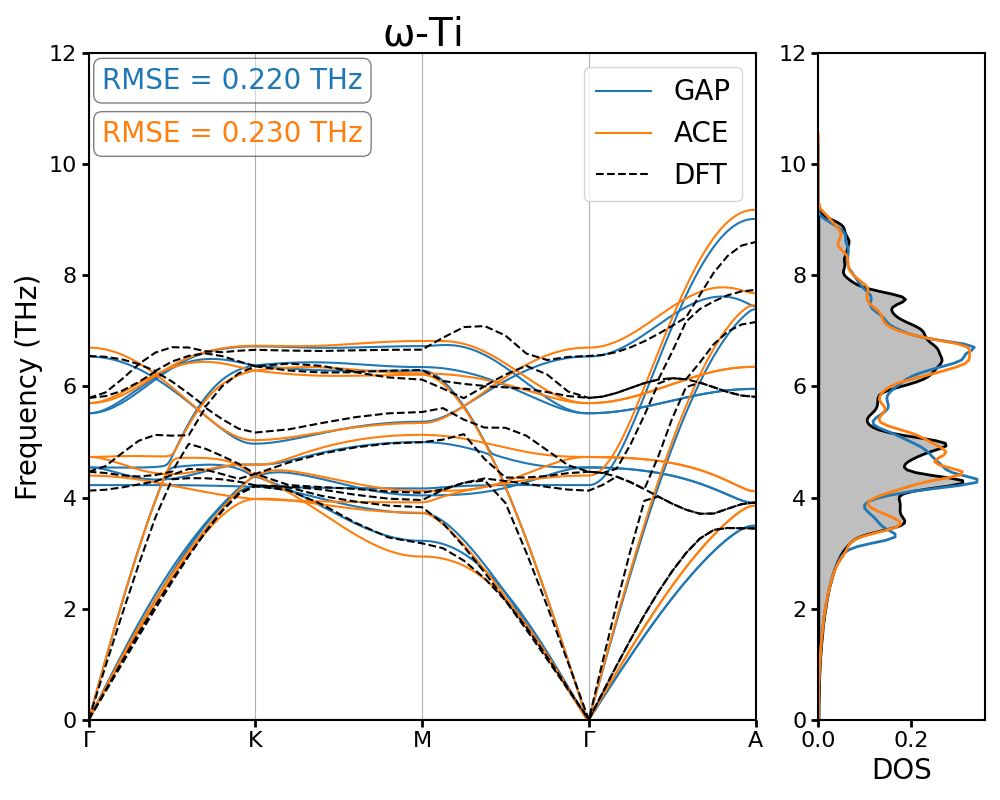}
    \caption[Phonons: $\omega$-Ti.]{\raggedright Phonon dispersion and density of states for $\omega$-Ti as calculated by the reference \ac{DFT} calculation alongside \ac{GAP} and \ac{ACE} surrogate models developed.}
    \label{fig:mTi_phonons_hex_Ti}
\end{figure}

\begin{figure}[h!tb]
    \centering
    \includegraphics[width=0.8\linewidth]{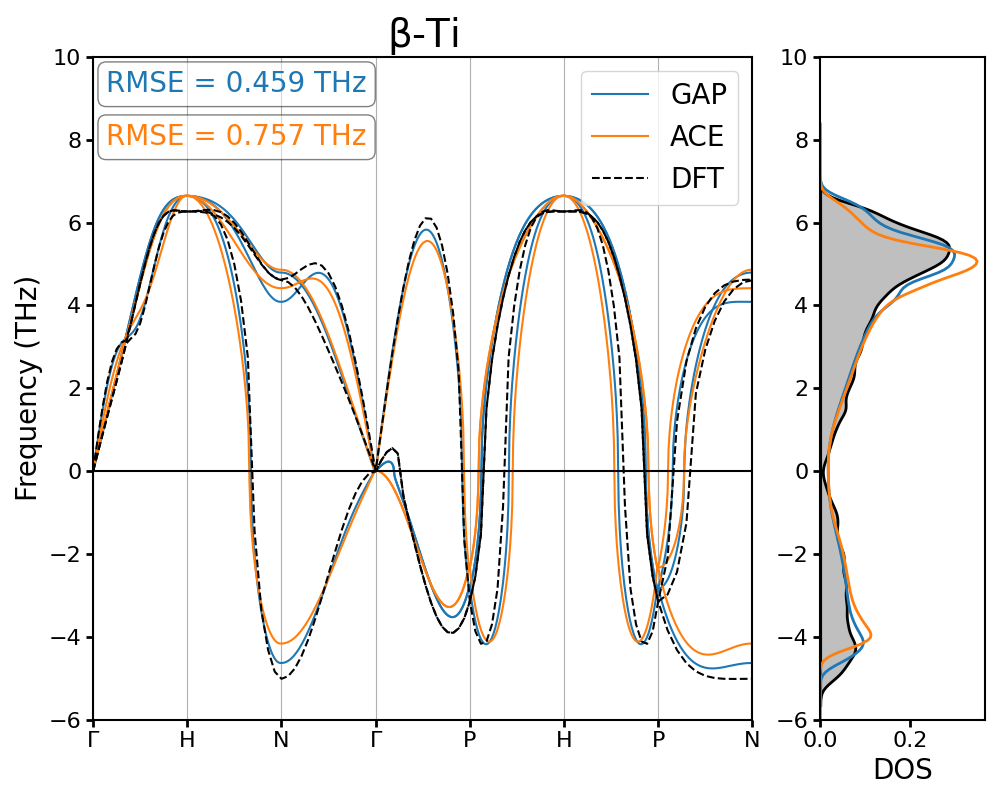}
    \caption[Phonons: $\beta$-Ti.]{\raggedright Phonon dispersion and density of states for $\beta$-Ti as calculated by the reference \ac{DFT} calculation alongside \ac{GAP} and \ac{ACE} surrogate models developed.}
    \label{fig:mTi_phonons_bcc_Ti}
\end{figure}

The phonon dispersions are shown along high-symmetry lines within the vibrational \ac{BZ}\cite{setyawan_high-throughput_2010} for our underlying \abinitio calculator and surrogate models in Figures \ref{fig:mTi_phonons_hcp_Ti},  \ref{fig:mTi_phonons_hex_Ti} and  \ref{fig:mTi_phonons_bcc_Ti} respectively.
For all crystalline phases, we found excellent agreement between the \ac{GAP} and \ac{ACE} potential with our DFT calculations, and this was achieved in an efficient manner, by using data that specifically targets the vibrational properties of each crystalline phase via the \ac{NDSC} method within our database building.
Our phonon dispersions are in excellent agreement with Hu \etal \cite{hu_theoretical_2010}, whilst some qualitative disagreement appears compared to the results presented by Nitol \etal \cite{nitol_machine_2022}\ in case of $\alpha$-Ti and $\beta$-Ti. \\

\subsection{Ti-6Al-4V}
\subsubsection{Fitting Machine Learned interatomic Potentials}\label{sec:ti64_MLIP_fitting}
We developed a series of multiphase potentials for the Ti-6Al-4V alloy.
A careful study of hyperparameters resulted in the final iteration of the \ac{GAP} model, which was constructed utilising 2-body kernels and many-body terms using the \verb|SOAPTurbo| descriptor.
The 2-body kernels utilised an inverse polynomial basis set with exponents -4, -8, -12 and -14 with spatial cutoff of 3.5~\AA{}, where 20 uniformly selected sparse points were used per elemental interaction type.
The \verb|SOAPTurbo| kernels consisted of 3563, 3700, and 2492 sparse points for Al, Ti and V centred environments, respectively.
The following \verb|SOAPTurbo| hyperparameters are utilised in the final iteration of our \ac{GAP} surrogate: $n_{\textrm{max}}$=8, $l_{\textrm{max}}$=8, $r_{\textrm{soft}}$=5.5 \AA, $r_{\textrm{hard}}$=6.0 \AA, $\sigma_{\parallel} = \sigma_{\perp}$=0.5 \AA, and $\zeta=6$.

The database of atomic configurations were generated using the procedure we outlined in Section~\ref{sec:generation_Ti64}, subsequently used to fit a \ac{GAP} surrogate model which was able to generate stable \ac{MD} trajectories across a broad range of thermodynamic conditions.
However, when the same database was used to fit an \ac{ACE} model, we observed that the V-V interaction was poorly characterised such that V mobility in molecular dynamics simulations was too great and non-physical interatomic distances were recorded.
To address this problem, we included additional configurations to capture V-V interactions.
In the configurations representative of Ti-6Al-4V via the stoichiometries $\alpha-\textrm{Ti}_{42}\textrm{Al}_{5}\textrm{V}_{7}$, $\beta-\textrm{Ti}_{19}\textrm{Al}_{3}\textrm{V}_{5}$, and $\omega-\textrm{Ti}_{26}\textrm{Al}_{4}\textrm{V}_{6}$, we fixed the positions of a pair of V-V nearest neighbours, and performed a geometry optimisation with \ac{DFT} relaxing all the other atoms.
We then displaced one of the V atoms such that the its distance to its nearest V neighbour was less than 1.6 \AA{}.
We also performed active learning, which included a set of configurations obtained from performing molecular dynamics on each crystal symmetry containing the stoichiometry $\textrm{Ti}_{8}\textrm{Al}_{1}\textrm{V}_{2}$, taking snapshots at 0.5 ps intervals, using the corresponding supercells:  $2\times2\times3$ ($\alpha$), $3\times3\times3$ ($\beta$), and $2\times2\times3$ ($\omega$).
The total size of the final iteration of the dataset was 8507 configurations with 147522 atomic environments.
The final version of the \ac{ACE} potentials presented here utilised $\nu=3$, for a \textit{total degree} of 15 across each correlation order, totalling 29,703 basis functions.
To keep our developed \acp{MLIP} commensurate, we also included this additional data in our final version of our \ac{GAP} surrogate that was not described in Section \ref{sec:ti64_database}. 
\begin{figure}[h!tb]
     \centering
     \begin{subfigure}[b]{\linewidth}
         \centering
         \includegraphics[width=\textwidth]{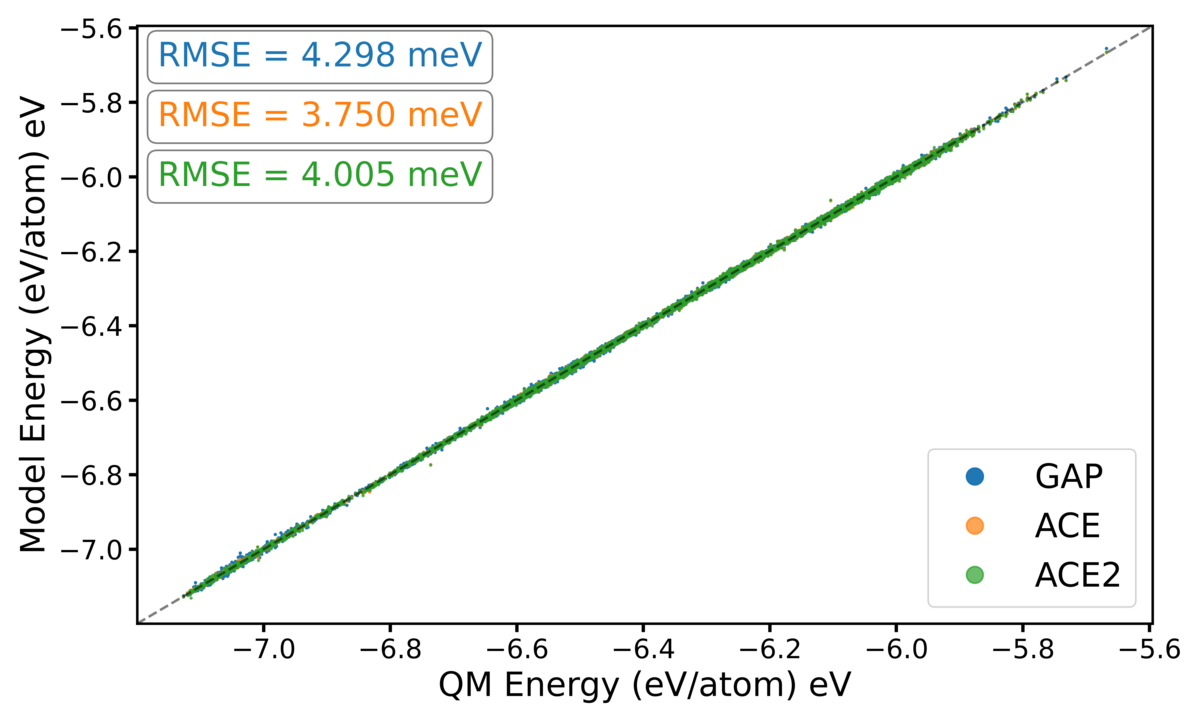}
     \end{subfigure}
     \begin{subfigure}[b]{\linewidth}
         \centering
         \includegraphics[width=\textwidth]{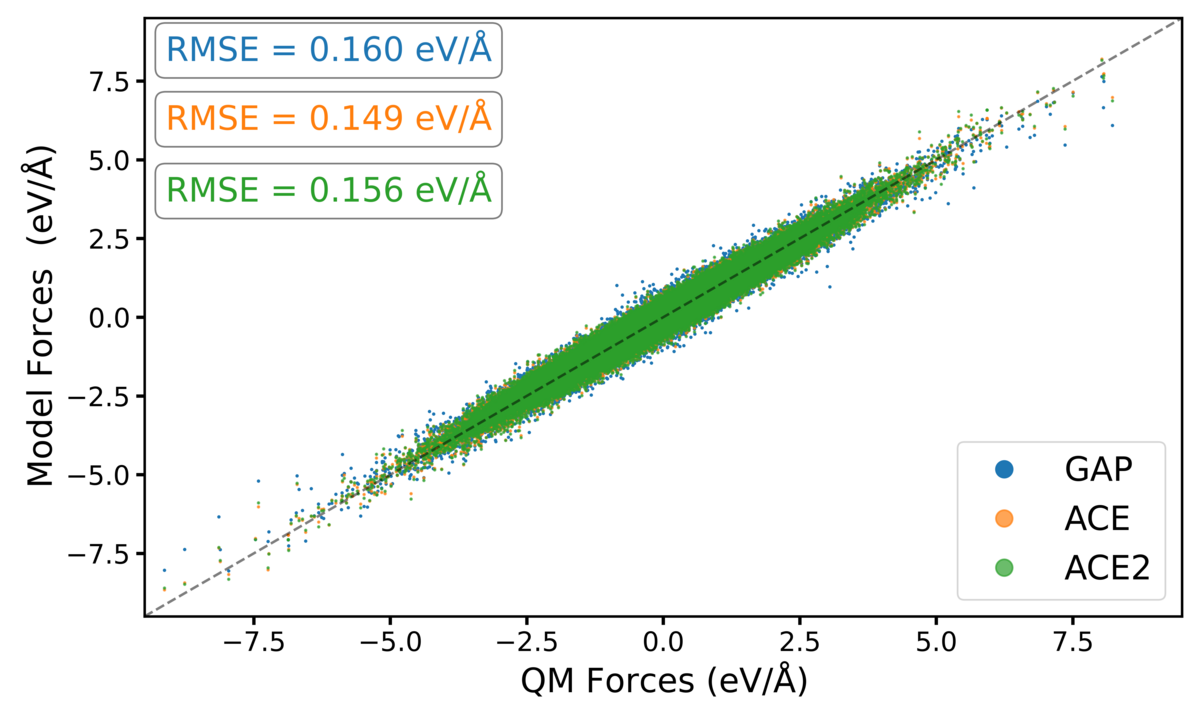}
     \end{subfigure}
     \begin{subfigure}[b]{\linewidth}
         \centering
         \includegraphics[width=\textwidth]{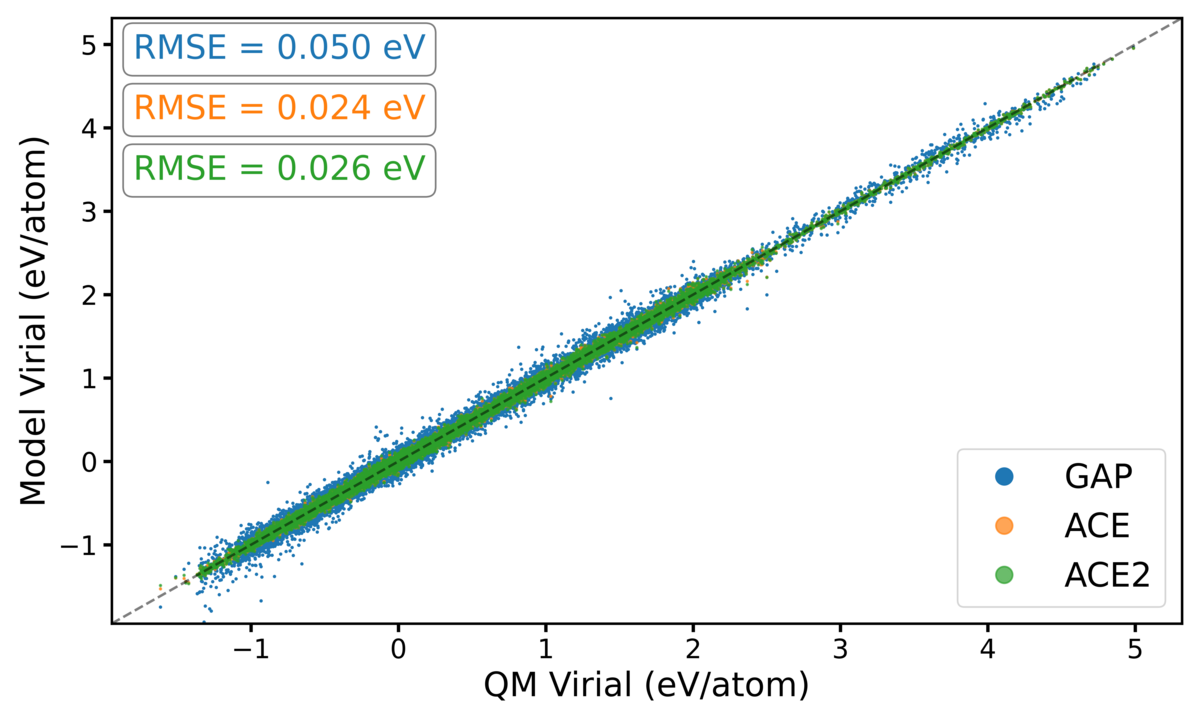}
     \end{subfigure}
        \caption{\raggedright Performance of surrogate models compared to training observables for each model in Ti-6Al-4V.}
        \label{fig:Ti64_RMSE_train}
\end{figure}

\begin{figure}[h!tb]
     \centering
     \begin{subfigure}[b]{\linewidth}
         \centering
         \includegraphics[width=\textwidth]{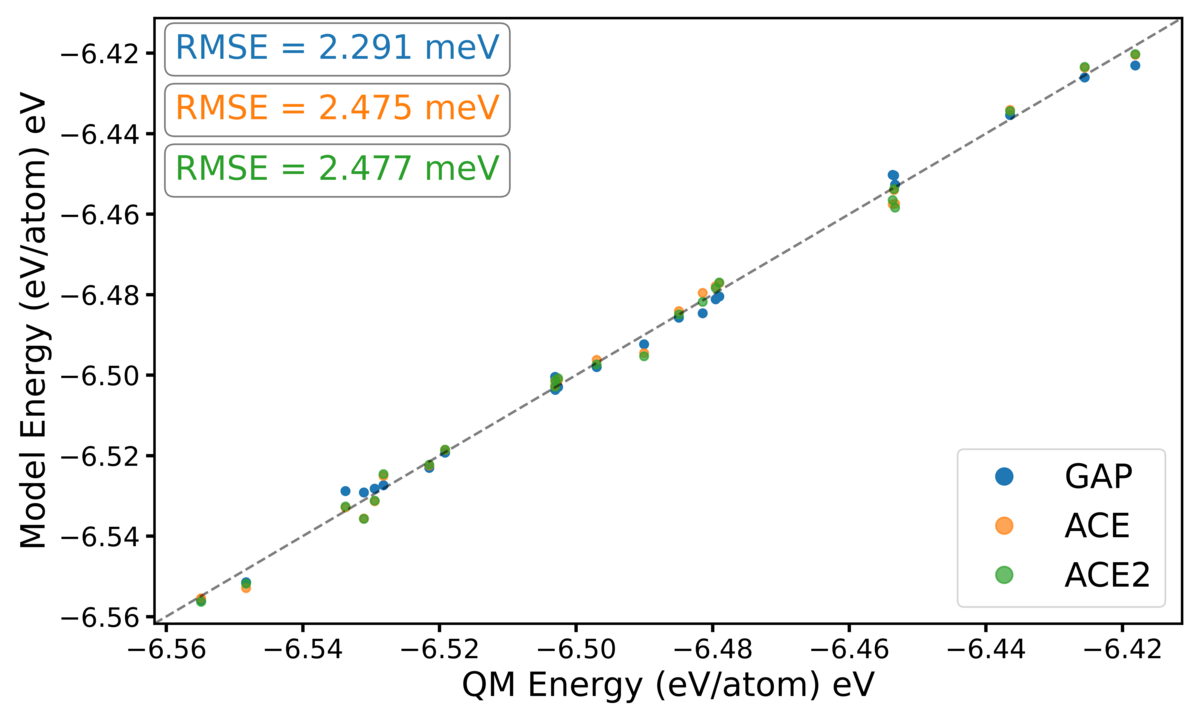}
     \end{subfigure}
     \begin{subfigure}[b]{\linewidth}
         \centering
         \includegraphics[width=\textwidth]{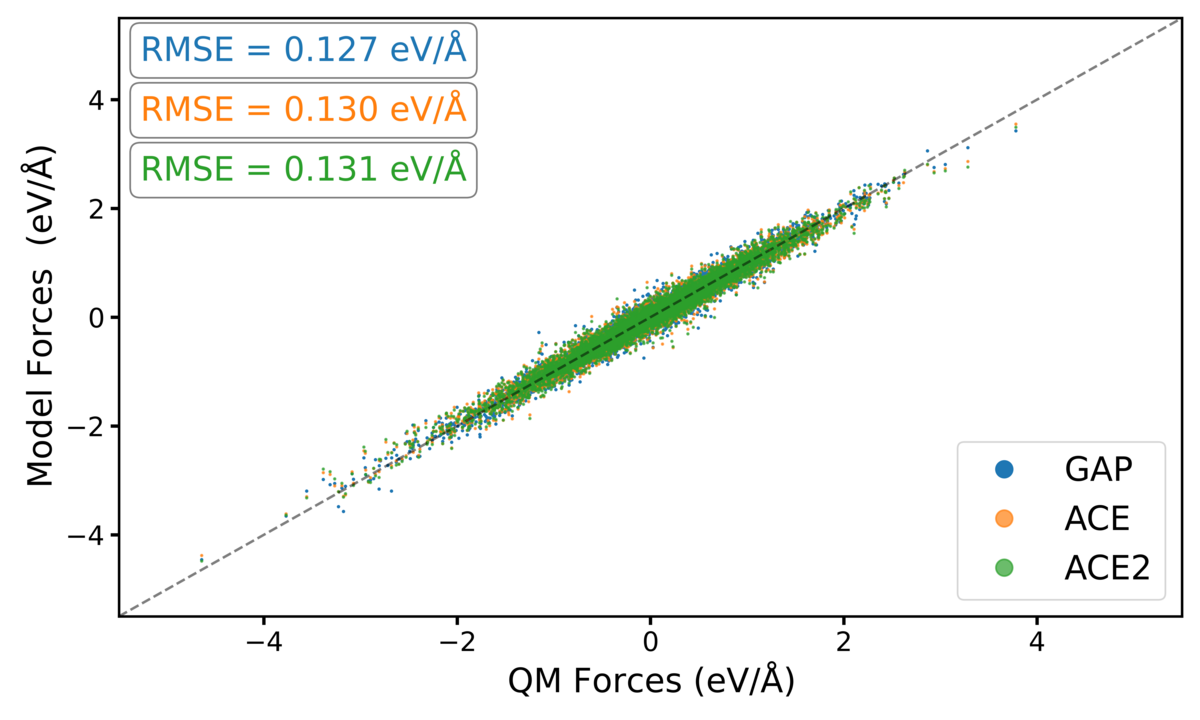}
     \end{subfigure}
     \begin{subfigure}[b]{\linewidth}
         \centering
         \includegraphics[width=\textwidth]{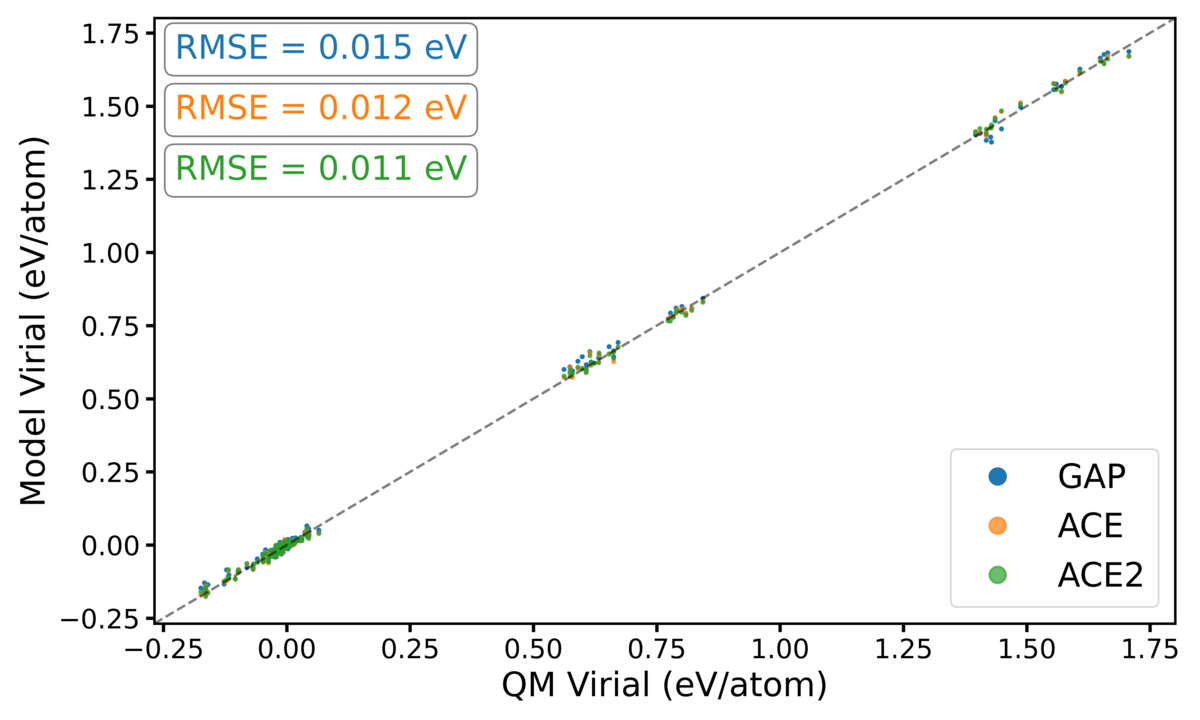}
     \end{subfigure}
        \caption{\raggedright Performance of surrogate models compared to observables in a validation dataset  for each model in Ti-6Al-4V.}
        \label{fig:Ti64_RMSE_validation}
\end{figure}

The \acp{MLIP} developed in this work are trained on the total configurational energy, atomic forces and virial stresses, and we evaluate the model performance of the final iteration of models developed on reproducing these quantities in the total training data in Figure \ref{fig:Ti64_RMSE_train}.
We compare each surrogate model's prediction with the \ac{DFT} prediction and compute the \ac{RMSE} for each quantity using the full dataset, which constituted 8507 \abinitio{ }calculations for a total of 502115 observables.
As indicated by the figure, both \ac{GAP} and \ac{ACE} surrogates reproduce the underpinning training data energy labels with accuracy of the order of meV/atom.
Compared to training results presented for our multiphase Ti potential in Chapter \ref{ch:Ti}, we note that the \acp{RMSE} of our multiphase Ti-6Al-4V surrogate models are comparable to the single element multiphase Ti models, despite the added complexity arising from chemical permutations.

To test the ability of the \ac{NDSC} approach to gather relevant configurations in a multi-elemental bulk phase, we evaluate our models against a validation dataset of configurations that represent the Ti-6Al-4V stoichiometry, where the details of its construction as discussed in Section \ref{sec:Ti64_DFT}.
The total validation dataset constitutes 23 configurations with 2604 atomic environments and 7973 observables.
In Figure \ref{fig:Ti64_RMSE_validation}, we present the performance of both surrogate models by comparing the predicted configurational energy, atomic forces and virial stresses in this validation set, and find that both models reproduce the underlying \abinitio{} calculations to similar accuracy of that of the training observables.

\begin{figure*}[h!tb]
     \centering
         \includegraphics[width=0.32\textwidth]{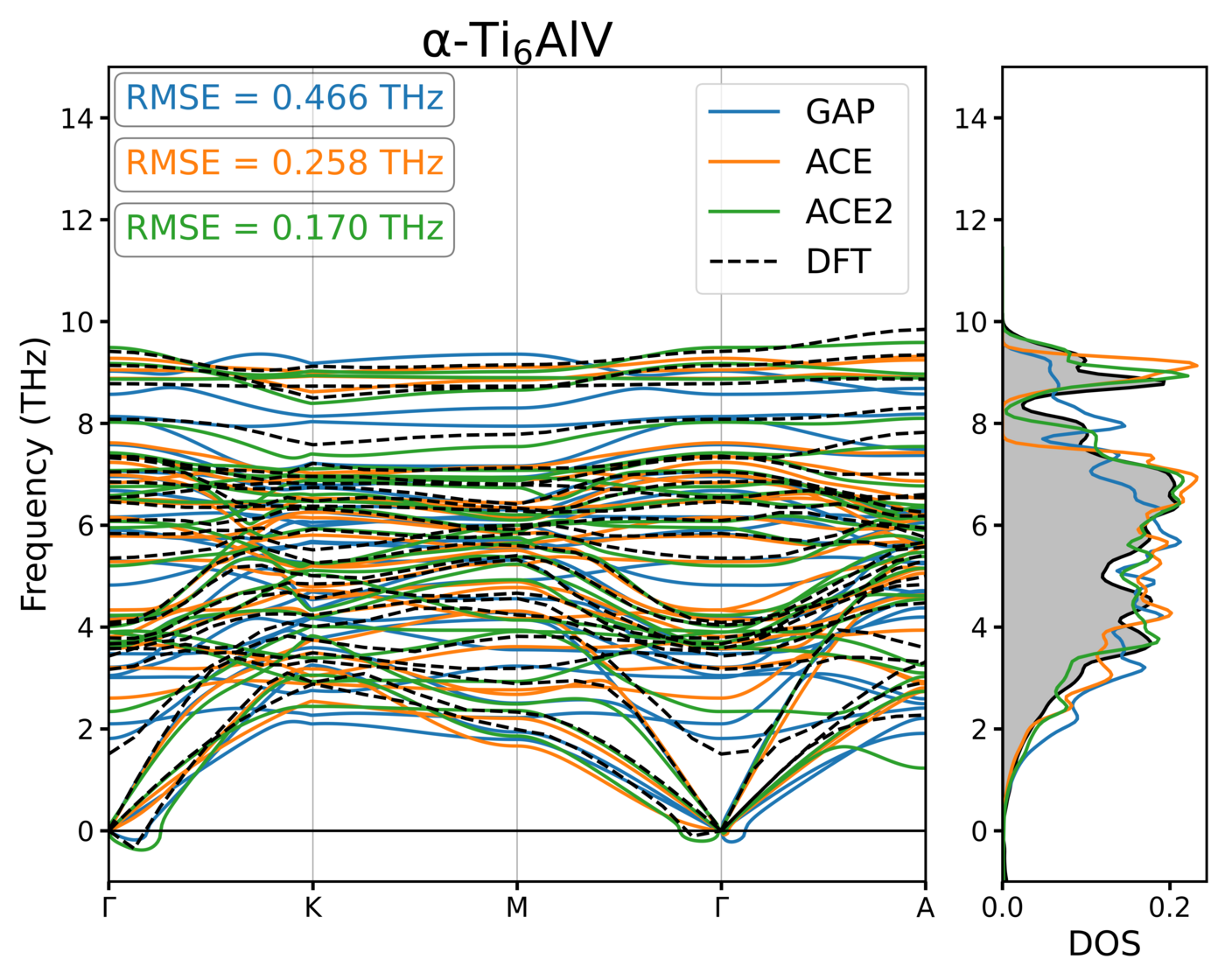}
         \includegraphics[width=0.32\textwidth]{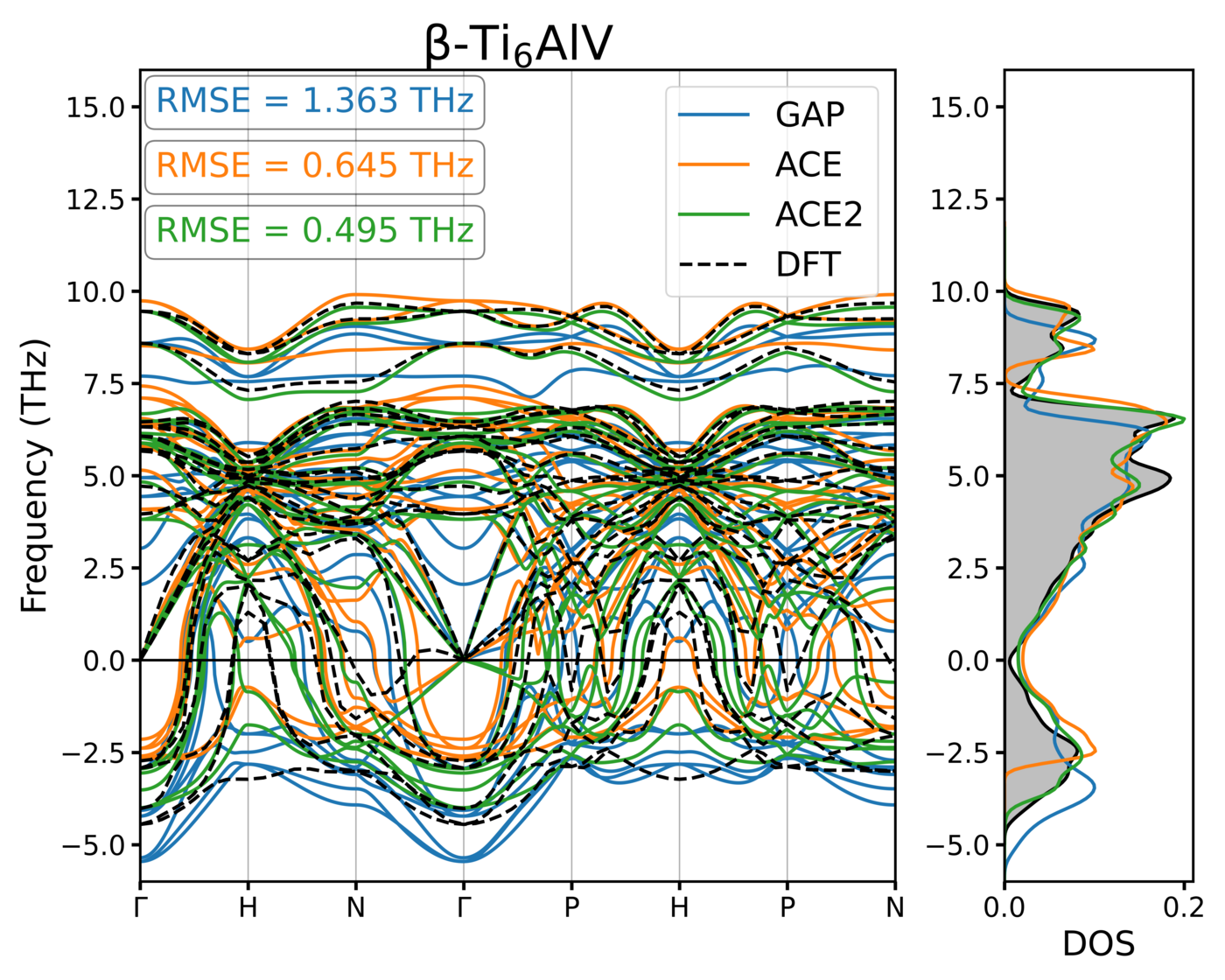}
         \includegraphics[width=0.32\textwidth]{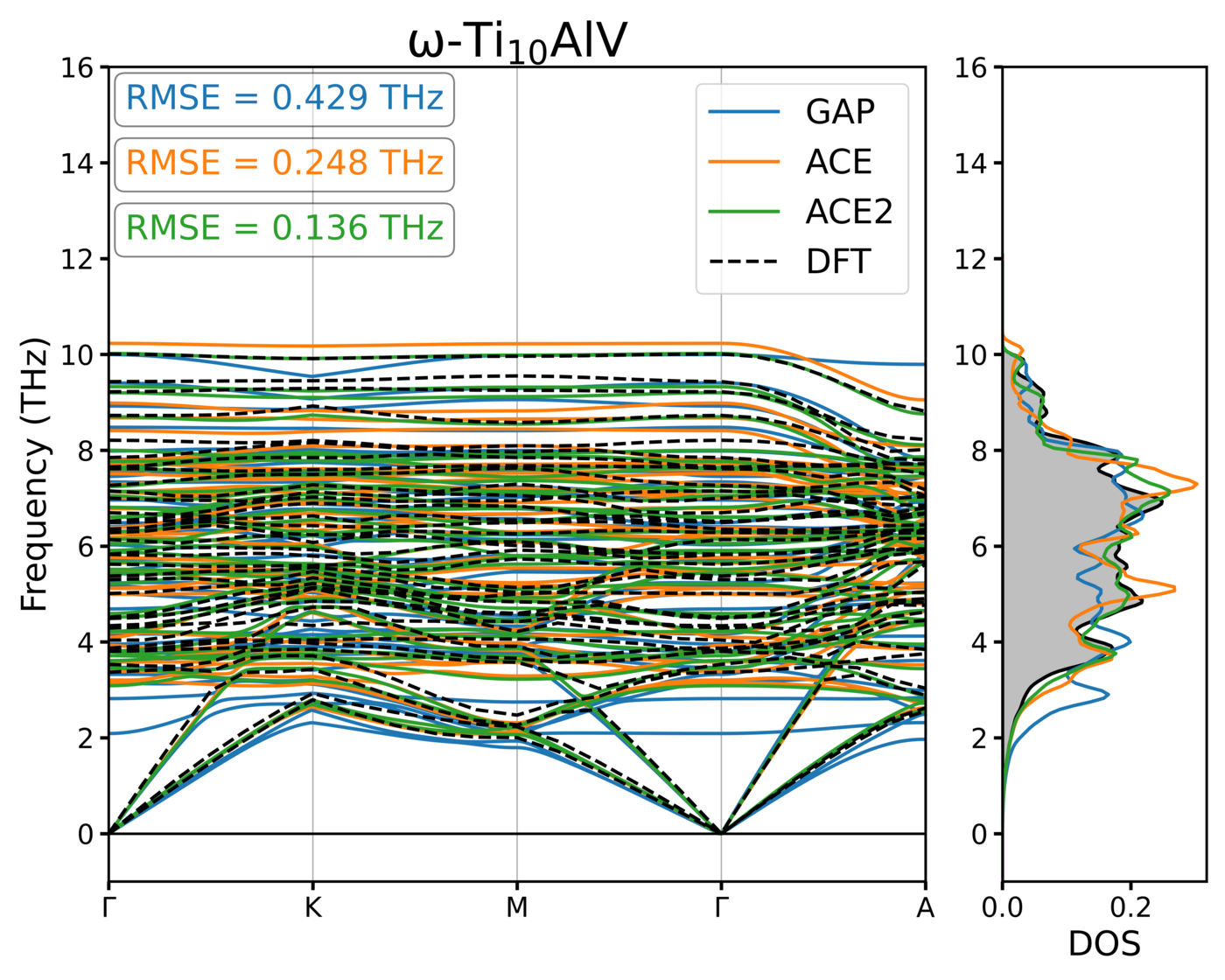}\\
         \includegraphics[width=0.32\textwidth]{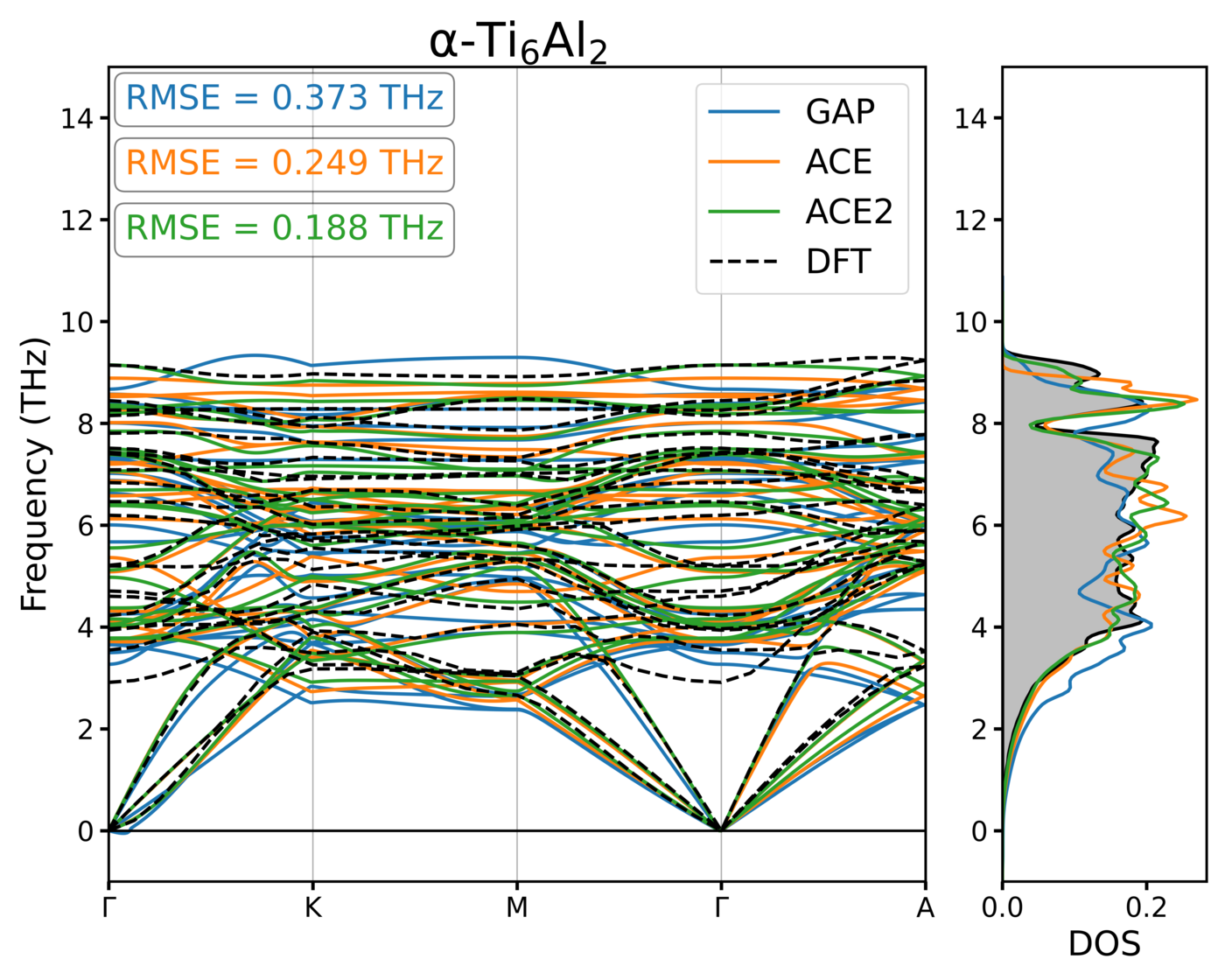}
         \includegraphics[width=0.32\textwidth]{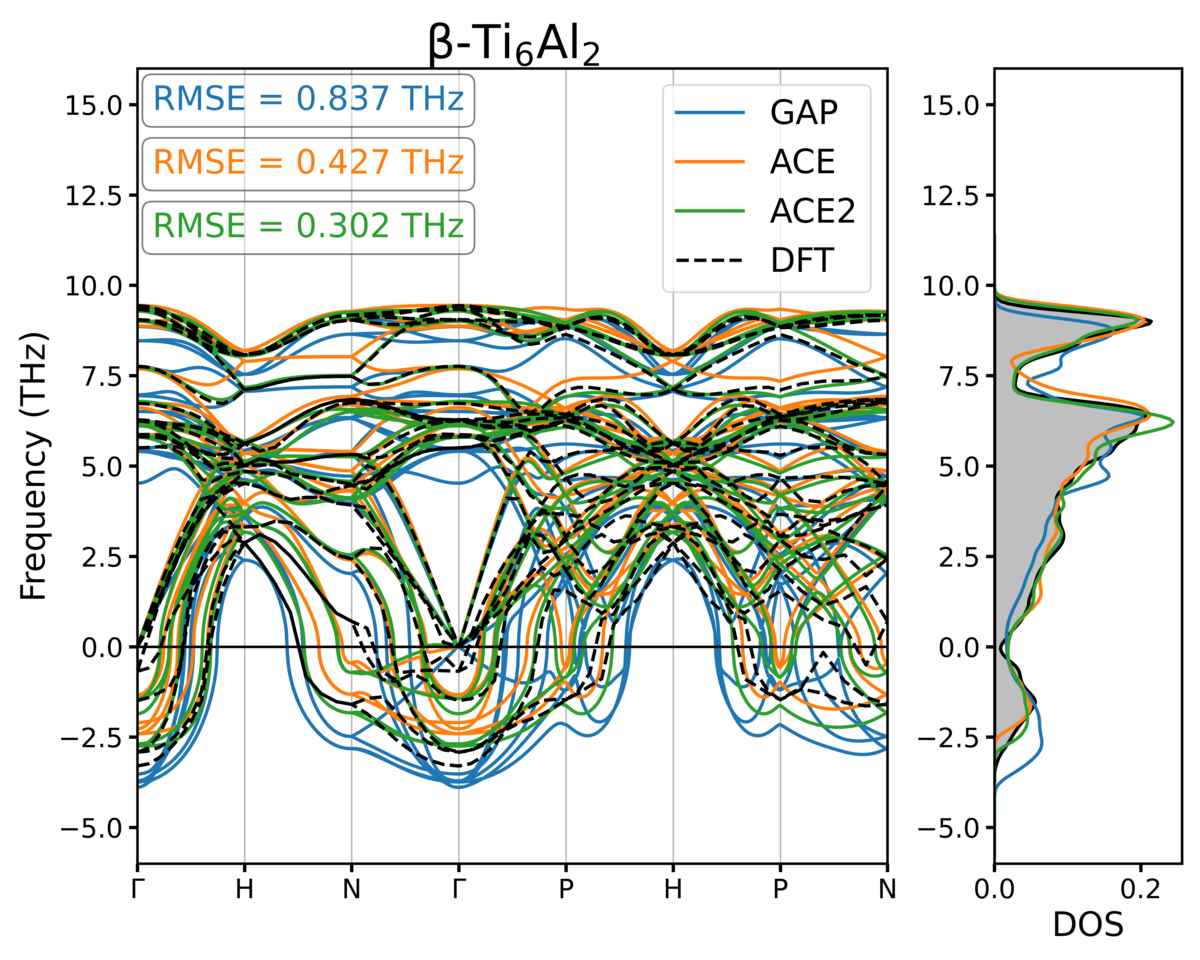}
         \includegraphics[width=0.32\textwidth]{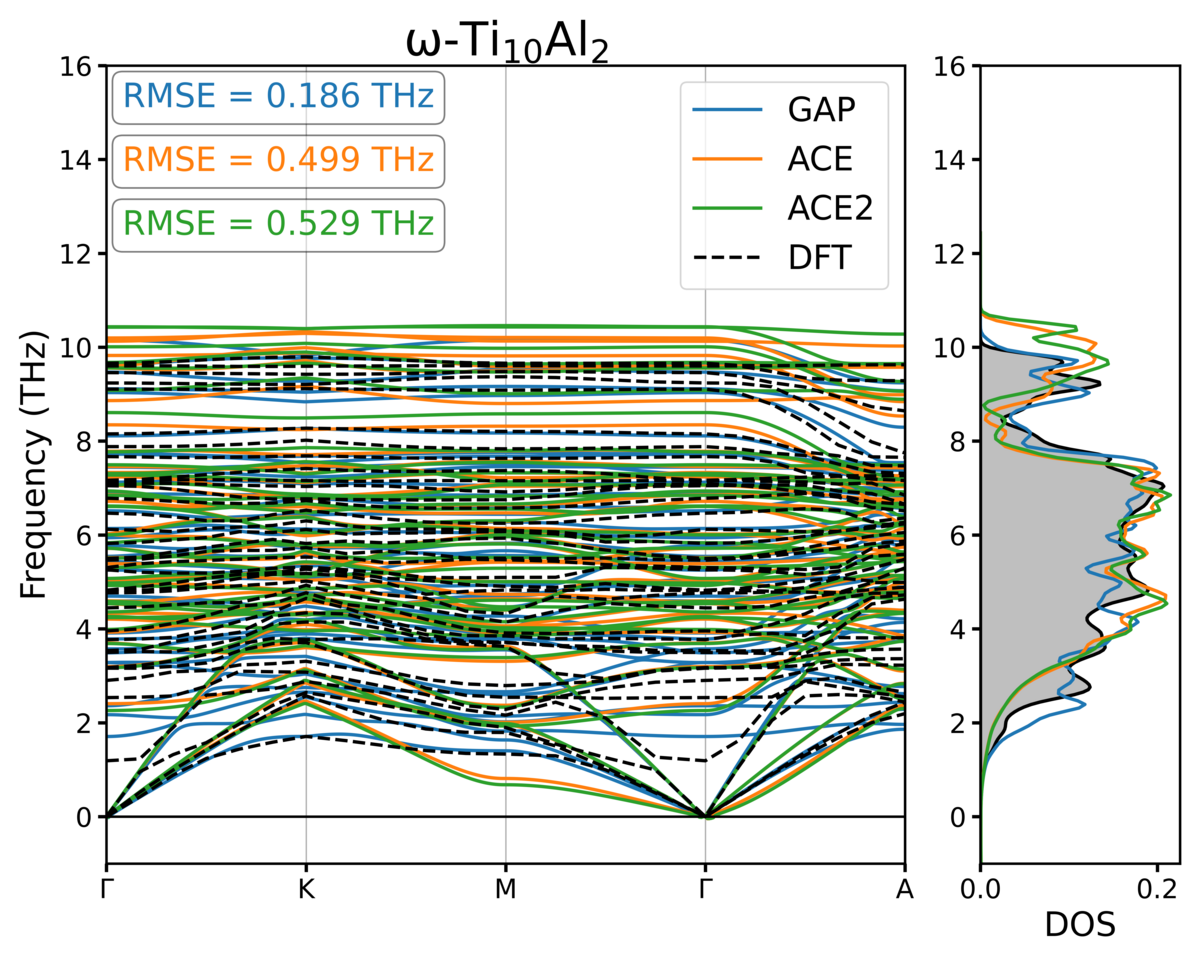}\\
         \includegraphics[width=0.32\textwidth]{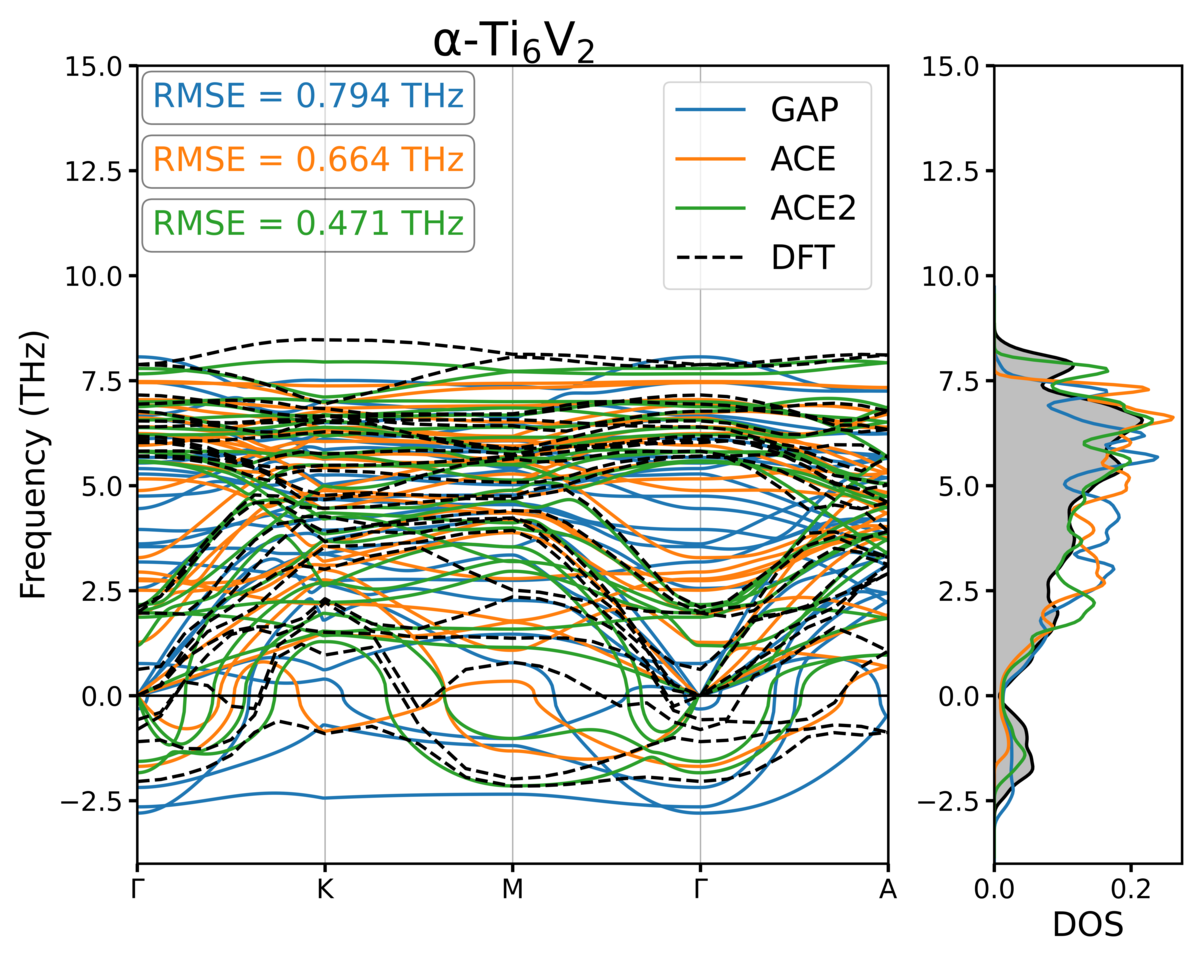}
         \includegraphics[width=0.32\textwidth]{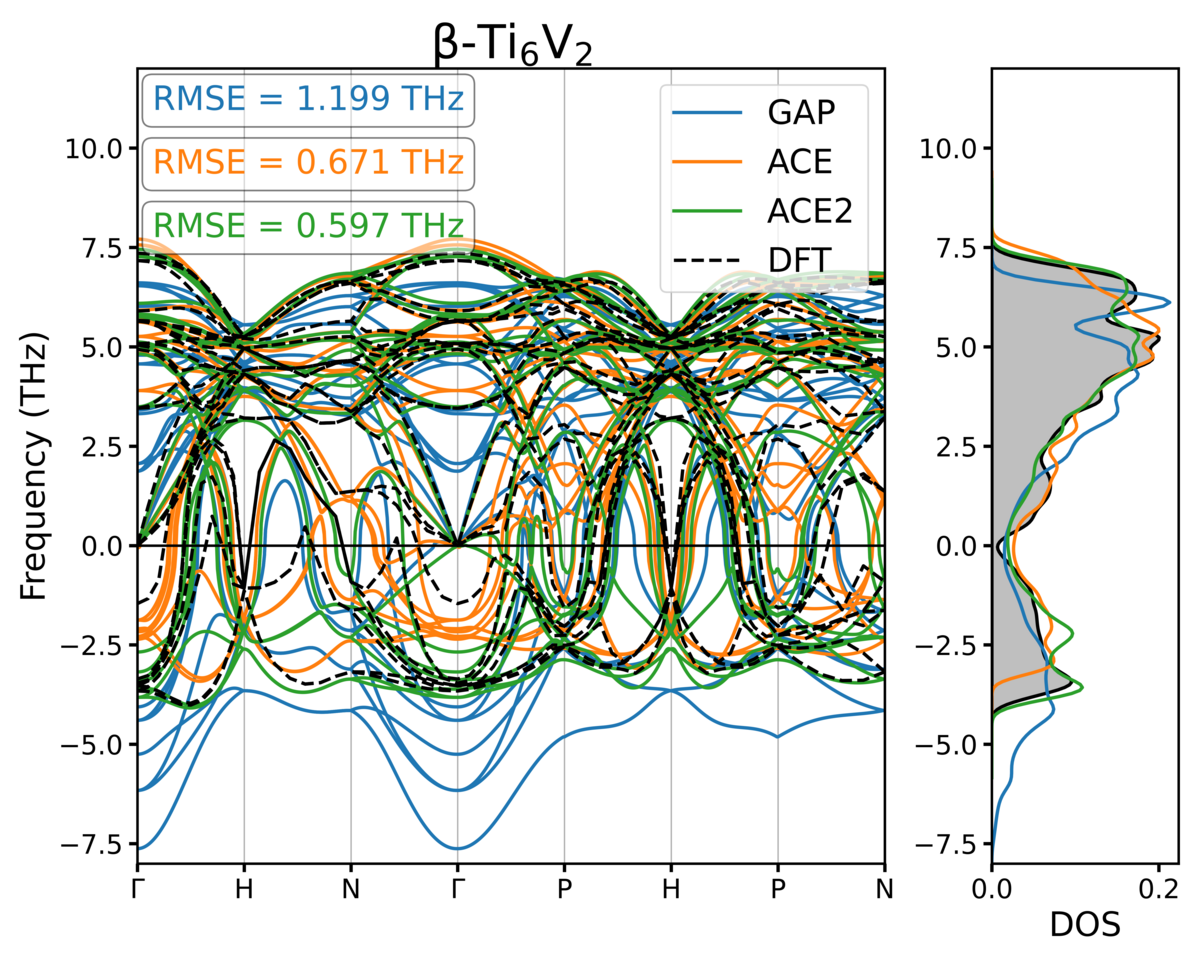}
         \includegraphics[width=0.32\textwidth]{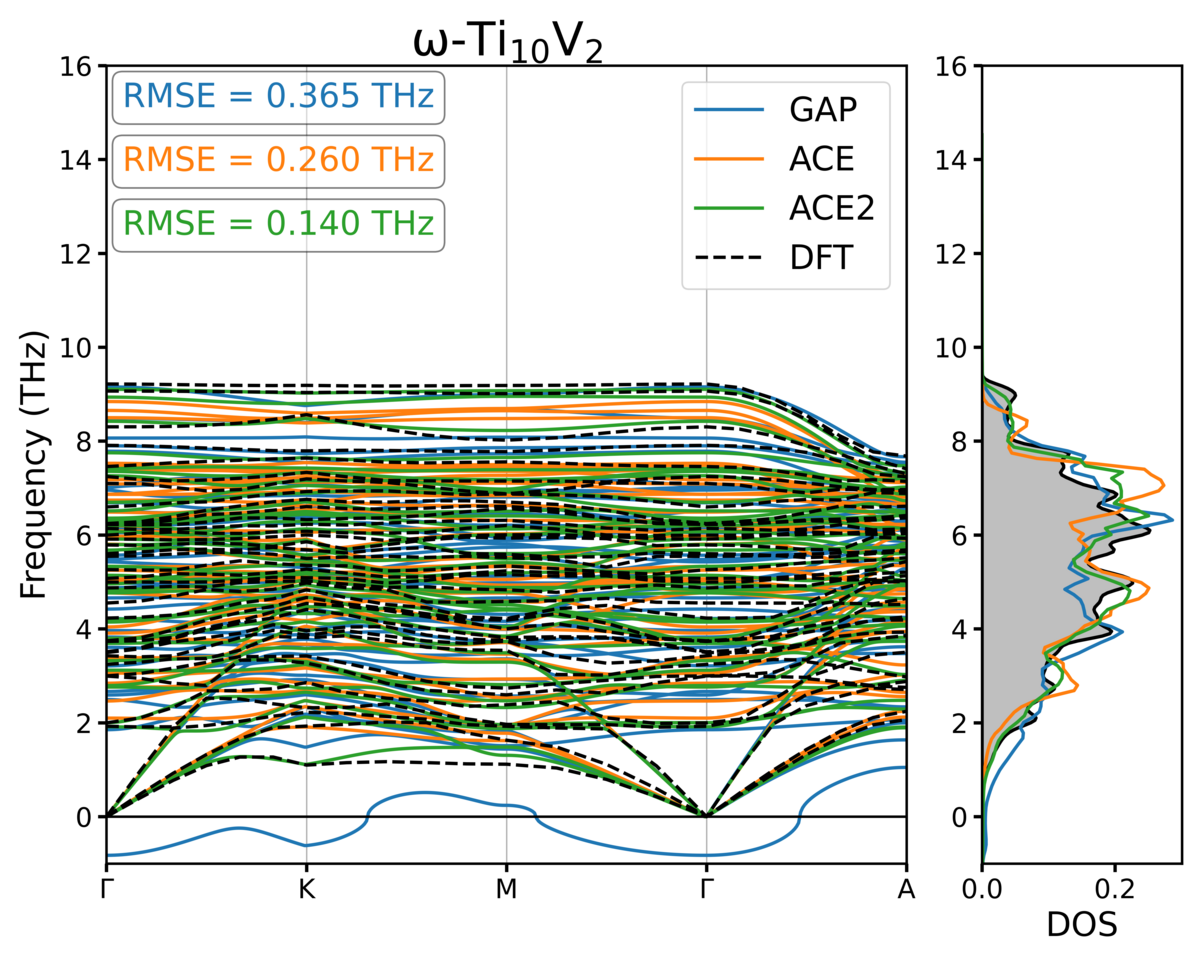}
        \caption{\raggedright Phonon dispersion and density of states as calculated with reference \ac{DFT} and developed \acp{MLIP} on the Ti-6Al-4V dataset. Minor alloying components (Al and V) are placed on nearest neighbouring sites.}
        \label{fig:phonon_all}
\end{figure*}

\subsubsection{Elastic Constants}

\begin{table}[h!tb]
    \centering
    \begin{tabular}{||l|c|c|c|c||}
    \hline
    $\alpha$-Ti-6Al-4V 0~K &  \hspace{0.2cm} DFT \hspace{0.2cm} &  \hspace{0.2cm} GAP \hspace{0.2cm} &  \hspace{0.2cm} ACE \hspace{0.2cm} &  \hspace{0.2cm} ACE2 \hspace{0.2cm} \\
    \hline
    Elastic constants (GPa): &&&&\\
    \hspace{0.2cm} \footnotesize{$C_{11}$} &198.3 & 197.7 & 199.4 & 199.65\\
    \hspace{0.2cm} \footnotesize{$C_{33}$} &198.2 & 221.3 & 208.4 & 209.2\\
    \hspace{0.2cm} \footnotesize{$C_{12}$} &67.2  & 69.2  & 65.4  & 66.6\\
    \hspace{0.2cm} \footnotesize{$C_{13}$} &73.8  & 61.0  & 70.5 & 68.1\\
    \hspace{0.2cm} \footnotesize{$C_{44}$} &48.9  & 41.4  & 49.6 & 49.0\\
    \hspace{0.2cm} \footnotesize{$C_{66}$} &65.6  & 64.2  & 67.0 & 66.5\\
    \hspace{0.2cm} \footnotesize{$B$}   &113.8 & 111.0 & 113.2 & 112.6\\
    Lattice parameters: &&&&\\
    \hspace{0.2cm} \footnotesize{$a$} (\AA)& 8.739 & 8.739  & 8.737 & 8.736\\
    \hspace{0.2cm} \footnotesize{$c$} (\AA)&   13.887& 13.888 & 13.885& 13.882\\
    \hspace{0.2cm} \footnotesize{$V_0$} (\AA$^3$/atom)& 17.01 & 17.01 & 17.00 & 16.99\\
    \hline
    \end{tabular}
    \caption[Elastic constants: $\alpha$-Ti-6Al-4V 0~K unrelaxed.]{\raggedright Unrelaxed elastic constants and lattice parameters for $\alpha$-Ti-6Al-4V at 0~K utilising a $3\times3\times3$ supercell.}
    \label{tab:Ti64_EC_alpha}
\end{table}

\begin{table}[h!tb]
    \centering
    \begin{tabular}{||l|c|c|c|c||}
    \hline
    $\beta$-Ti-6Al-4V 0~K &  \hspace{0.2cm} DFT \hspace{0.2cm} &  \hspace{0.2cm} GAP \hspace{0.2cm} &  \hspace{0.2cm} ACE \hspace{0.2cm} &  \hspace{0.2cm} ACE2 \hspace{0.2cm} \\
    \hline
    Elastic constants (GPa): &&&&\\
    \hspace{0.2cm} \footnotesize{$C_{11}$} &153.0 & 160.2 & 154.2 & 156.5\\
    \hspace{0.2cm} \footnotesize{$C_{12}$} &90.7  & 81.6  & 84.1  & 84.4 \\
    \hspace{0.2cm} \footnotesize{$C_{44}$} &60.9  & 74.2  & 61.2  & 63.1 \\
    \hspace{0.2cm} \footnotesize{$B$}   &111.5 & 107.8 & 107.5 & 108.5\\
    Lattice parameters: &&&&\\
    \hspace{0.2cm} \footnotesize{$a$} (\AA)& 11.231 & 11.231  & 11.237 & 11.240\\
    \hspace{0.2cm} \footnotesize{$V_0$} (\AA$^3$/atom)& 17.04 & 17.04 & 17.07 & 17.08\\
    \hline
    \end{tabular}
    \caption[Elastic constants: $\beta$-Ti-6Al-4V 0~K unrelaxed.]{\raggedright Unrelaxed elastic constants and lattice parameters for $\beta$-Ti-6Al-4V at 0~K utilising a $4\times4\times4$ supercell.}
    \label{tab:Ti64_EC_beta}
\end{table}

\begin{table}[h!tb]
    \centering
    \begin{tabular}{||l|c|c|c|c||}
    \hline
    $\omega$-Ti-6Al-4V 0~K &  \hspace{0.2cm} DFT \hspace{0.2cm} &  \hspace{0.2cm} GAP \hspace{0.2cm} &  \hspace{0.2cm} ACE \hspace{0.2cm} &  \hspace{0.2cm} ACE2 \hspace{0.2cm} \\
    \hline
    Elastic constants (GPa): &&&&\\
    \hspace{0.2cm} \footnotesize{$C_{11}$} & 195.9 &198.5  &190.7  & 190.8\\
    \hspace{0.2cm} \footnotesize{$C_{33}$} & 229.1 & 254.4 &215.6  &218.5 \\
    \hspace{0.2cm} \footnotesize{$C_{12}$} & 84.5 &  91.5 & 84.1 & 83.1\\
    \hspace{0.2cm} \footnotesize{$C_{13}$} & 57.1 &  43.3 & 56.3 & 55.8\\
    \hspace{0.2cm} \footnotesize{$C_{44}$} & 51.0 &  46.5 & 51.3 & 50.6\\
    \hspace{0.2cm} \footnotesize{$C_{66}$} & 55.7 &  53.5 & 53.3 & 53.9\\
    \hspace{0.2cm} \footnotesize{$B$}   & 113.1& 111.9 & 110.1 &110.0 \\
    Lattice parameters: &&&&\\
    \hspace{0.2cm} \footnotesize{$a$} (\AA)& 13.670 & 13.678  & 13.673 & 13.678 \\
    \hspace{0.2cm} \footnotesize{$c$} (\AA)&  8.465 & 8.469   & 8.466  & 8.469\\
    \hspace{0.2cm} \footnotesize{$V_0$} (\AA$^3$/atom)& 16.91 & 16.94 & 16.92  & 16.94\\
    \hline
    \end{tabular}
    \caption[Elastic constants: $\omega$-Ti-6Al-4V 0K unrelaxed.]{\raggedright Unrelaxed elastic constants and lattice parameters for $\omega$-Ti-6Al-4V at 0~K utilising a $3\times3\times3$ supercell.}
    \label{tab:Ti64_EC_omega}
\end{table}

We computed the elastic properties of the cells representative of the Ti-6Al-4V stoichiometry.
In these benchmarks, as described in Section \ref{sec:Ti64_DFT}, we utilise examples of each crystalline symmetry of large supercells containing 54, 64, and 81 ($\alpha$, $\beta$ and $\omega$, respectively) atoms. When computing the elastic constants using finite differences, prior to the deforming the lattice, geometry relaxation using the appropriate \ac{MLIP} model was carried out to provide the ground state lattice parameters and atomic positions. When calculating macroscopic elasticity properties using numerical differentiation, atomic positions of each finite strain configuration should be relaxed.
However, this approach would have incurred a significant additional computational cost when calculating the reference values using \ac{DFT}, due to the large number of atoms in the configurations.
As the main purpose of this calculation is to provide benchmark values, to be used in validating \ac{MLIP} models, we decided to keep the atomic positions unrelaxed.
As a result, our reported elastic constants are not commensurate with experimental observations and therefore not relevant in characterising the macroscopic properties of Ti-6Al-4V, only serve to provide insight on the quality of interpolation by the \acp{MLIP} developed in this work.
We refer to these elastic constants as {\em unrelaxed}, which are presented in Tables~\ref{tab:Ti64_EC_alpha}, \ref{tab:Ti64_EC_beta} and \ref{tab:Ti64_EC_omega}.


Using a fast surrogate model allows us to predict elastic constants at finite temperatures, which can be related to experimental observables.
We used our \ac{ACE} surrogate model to run \ac{MD} simulations in the canonical ensemble and computed the elastic constants from the fluctuation of the stress tensor elements\cite{ray_statistical_1984, zhen_deformationfluctuation_2012,clavier_computation_2017}.
We initialised a series of supercells with the orthorhombic versions of the $\alpha$-Ti symmetry with $13\times8\times9$ repeating units, randomly replacing Ti with Al and V to construct the Ti-6Al-4V stoichiometry.
The \ac{LAMMPS} package was used to propagate the dynamics and to monitor stress fluctuations for different strain patterns via the computation of the Born matrix.
We compare our results to single crystal \cite{heldmann_diffraction-based_2019} and polycrystalline \cite{STAPLETON20086186} experiments at room temperature alongside a theoretical result at the \ac{GGA} \ac{DFT} level of theory \cite{guler_first-principles_2021} in Table~\ref{tab:Ti64_EC_alpha_300K}. 



\begin{table}[h!tb]
    \centering
    \begin{tabular}{||l|c|c|c|c||}
    \hline
    $\alpha$-Ti-6Al-4V  &  \hspace{0.2cm} ACE \hspace{0.2cm} &  \hspace{0.2cm} Ex$_1$ \cite{heldmann_diffraction-based_2019} \hspace{0.2cm} & \hspace{0.2cm} Ex$_2$ \cite{STAPLETON20086186} \hspace{0.2cm} &  \hspace{0.2cm} DFT \cite{guler_first-principles_2021} \hspace{0.2cm}  \\
    300~K & &  &  &  \\
    \hline
    E.C. (GPa): &&&&\\
    \hspace{0.2cm} \footnotesize{$C_{11}$} &            158.5 (3.1)  & 168.0 & 143.0 & 153.2  \\
    \hspace{0.2cm} \footnotesize{$C_{12}$} &            89.8 (3.3)   & 108.0 & 110.0 & 55.1 \\
    \hspace{0.2cm} \footnotesize{$C_{13}$} &            71.4 (0.7)   & 39.0  & 90.0  & 48.6 \\
    \hspace{0.2cm} \footnotesize{$C_{33}$} &            189.2 (1.0)  & 144.0 & 177.0 & 157.2  \\
    \hspace{0.2cm} \footnotesize{$C_{44}$} &            43.3 (0.5)   & 44.0  & 40.0  & 37.1 \\
    \hline
    \end{tabular}
    \caption[Elastic constants: $\alpha$-Ti-6Al-4V 300~K.]{ \raggedright Finite temperature elastic constants and cell parameters for $\alpha$-Ti-6Al-4V at 300~K compared to literature values. The column Ex$_1$ corresponds to a single crystal experiment\cite{heldmann_diffraction-based_2019}, whereas Ex$_2$ shows elastic constants determined on polycrystalline samples\cite{STAPLETON20086186}. Predictions using \ac{DFT} are also presented\cite{guler_first-principles_2021}.}
    \label{tab:Ti64_EC_alpha_300K}
\end{table}

Our elastic constants, as predicted by our surrogate \ac{ACE} model, are found to be consistently between the single and polycrystalline experimental results, however, our model tends to over-predict the stiffness with respect to strains in the $\epsilon_{33}$ direction.
We also provide uncertainty estimates on our results that arise from different local permutations of the minor alloying components, and observe that the local ordering in a single crystal has a negligible effect on elastic constants.
At 0~K, we observed similar results in $\alpha$-Ti-6Al-4V with the finite differences method, where we computed 20 different realisations for supercells used previously to benchmark our \ac{MLIP} against our unrelaxed reference \ac{DFT} elastic constants, however, this time allowing for internal relaxation of atomic positions. We similarly computed the 0~K $\beta$-Ti-6Al-4V elastic constants for many random permutations, however, noting that due to the dynamic instability of the $\beta$ phase, we do not relax atomic positions, and find that local ordering contributes variability less than 5 GPa  to the elastic constants in our \ac{ACE} model. \\


\subsubsection{Vibrational Properties}
To evaluate how accurately the \ac{FCM} are reproduced by the \acp{MLIP} developed on the Ti-6Al-4V dataset, we compute phonon dispersions and density of states for a series of configurations that are tractable to high-symmetry points beyond $\Gamma$-point predictions with our \abinitio{} reference method.
Due to the small size of the unit cells we used in this benchmark, we note that the concentration of the minor alloying components Al and V, are consequently higher than in the Ti-6Al-4V alloy.
We considered configurations where the minor alloying components were nearest neighbours in the crystalline lattice for each  symmetry.
We note that Al-Al, Al-V and V-V interactions are only sparsely represented in the training database, therefore our benchmarks can be regarded as a stringent tests of extrapolative behaviour of the models.
We also compute the \ac{FCM} as predicted by \acp{MLIP} with the finite difference method using \verb_phonopy_ and from the \ac{FCM} we determine the phonon dispersion and density of state relations.
We show the phonon dispersions and density of states in Figure \ref{fig:phonon_all}.
Despite no training data point was specifically crafted to represent these configurations, we observed generally good agreement in most cases between the \ac{DFT} reference and our \ac{MLIP} models.
However, discrepancies in case of some of the phonon dispersion benchmark tests remain present.
In order to quantify whether inaccuracies in the surrogate models are due to insufficient training data or inadequate choice hyperparameters, such as the spatial cutoff distance or the body order representation, we carried out further numerical experiments.



\begin{figure}[h!tb]
    \centering
    \includegraphics[width=0.49\linewidth]{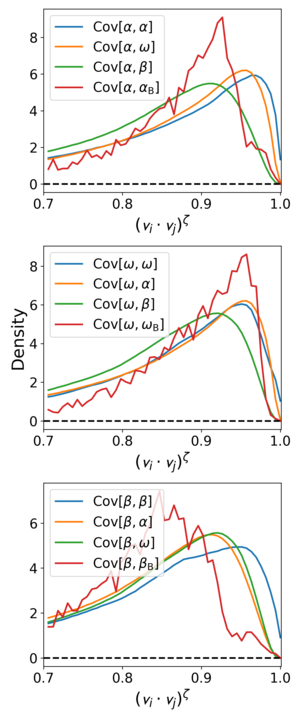}
    \includegraphics[width=0.49\linewidth]{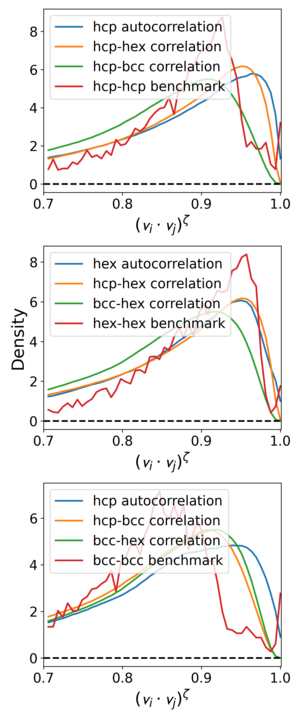}
    \caption[Covariance between phonon benchmarks and crystalline data.]{\raggedright Histogram of \ac{SOAP} similarity values.
    The left column represents the original database and the right column represents the database augmented with targeted data points.
    The three rows represent subsets of the training databases containing $\alpha$, $\beta$ and $\omega$ phase configurations compared to atomic configurations in each phase (blue, orange and green lines) as well as the unit cells used in the phonon benchmarks (red lines). Blue lines represent the similarities of atomic environments belonging to the same phase.}
    \label{fig:Ti64_correlations}
\end{figure}

We hypothesised that augmenting the original training data with data points specifically representing configurations that contain Al-Al, Al-V and V-V atom pairs at nearest neighbour crystalline sites would improve the accuracy of the predicted vibrational properties, if the reason for inaccuracies are due to inadequate data coverage.

Alongside all of our \ac{ACE} models we also considered a second variant of our \ac{ACE} model, labelled ACE2, which includes additional data
We generated \acp{NDSC} commensurate with a phonon grid sampling of $2\times2\times2$ for each phonon benchmark configuration, where atomic positions were displaced according to a normal distribution of standard deviation 0.05~\AA{}, to generate a total of 6 examples for every \ac{NDSC} of each phonon benchmark configuration.
This additional data constituted 374 \ac{DFT} calculations with 6500 atomic environments.
A new \ac{ACE} model was trained, appending the targeted data to the original dataset, with the weights on these targeted observables set as: $\boldsymbol{w}_E$=30, $\boldsymbol{w}_F$=25, and $\boldsymbol{w}_V$=5.
The ACE2 model may be interpreted as a best possible improvement at a given surrogate model hyperparameter set, and serves as a metric to understand the quality of the original \ac{ACE} which had no data that explicitly targeted these types of benchmarks.
The phonon dispersion \acp{RMSE} across all systems with the targeted data ACE2 model shows an average improvement of 21\%, which is similar to the improvement seen across phonon dispersion relations of \ac{ACE} over the \ac{GAP} model.

To provide further confirmation to our hypothesis, we analysed the similarity of atomic environments found in the original and the extended datasets, and comparing them to the atomic environments found in the configurations used for the phonon benchmark study.
We calculated the similarity, or covariance values, $C_{ij}$ of two atomic environment $i$ and and $j$ using their \ac{SOAP} descriptors $\mathbf{v}_i$ and $\mathbf{v}_j$ as
\begin{equation*}
    C_{ij} = (\mathbf{v}_i \cdot \mathbf{v}_j)^{\zeta}
\end{equation*}
using $\zeta = 4$.
Using this measure of similarity, $C_{ij} = 1$ corresponds to identical atomic environments and lower values signify different atomic environments.

All pairwise similarity values were calculated, and their histograms are presented in Figure~\ref{fig:Ti64_correlations}.
We grouped the values such that we present the similarities of environments found in the configurations used for the phonon benchmarks and those in the training databases representing the $\alpha$, $\beta$ and $\omega$ phases.
The histograms emphasise that atomic environments in the $\alpha$ and $\omega$ phases are more similar to each other than to those in the $\beta$ phase.
It is also revealed that in the original database the atomic environments typical for the phonon benchmark configurations are under-represented, which is remedied when augmenting the dataset using further \ac{NDSC} data points.

Given that our phonon benchmark tests are only indicative of the extrapolative behaviour of the generated \ac{MLIP} models, we conclude that the Ti-6Al-4V dataset is sufficient to produce \acp{MLIP} that accurately characterise the vibrational properties of the Ti-6Al-4V alloy where the concentrations of the minority components closely reflect those of the real material.\\

\section{Conclusions}
Two \abinitio{} datasets have been constructed for the purpose of \ac{MLIP} development for Ti and its technologically relevant Ti-6Al-4V alloy.
Each of these datasets was constructed by considering the experimentally observed condensed phases of Ti and Ti-6Al-4V below 30 GPa, respectively.
For both datasets, we utilise the \ac{NDSC} method as a strategy for accurately sampling the vibrational Brillouin zone of the crystalline systems.
In the case of Ti-6Al-4V, we have extended the \ac{NDSC} method as a data reduction strategy for the sampling of substitutional disorder within an atomic configuration.

We have fitted \acp{MLIP} using our datasets and tested if our \acp{MLIP} can accurately interpolate the Born-Oppenheimer \ac{PES}.
Focussing on structural dynamical properties, we have constructed validation tests based on our reference \ac{DFT} method, where our surrogate models showed excellent agreement with the reference method.

In the case of Ti-6Al-4V, we have also demonstrated the effectiveness of the data reduction strategy of the \ac{NDSC} method for sampling substitutional disorder in \ac{MLIP} development by showing the transferability of the developed \acp{MLIP} to configurations representing larger unit cells.  \\


\section{Data availability}
We make available the databases and presented models available in the dedicated repository: \url{https://zenodo.org/records/14244105}.

\section{Acknowledgments}
We acknowledge support from the NOMAD Centre of Excellence, funded by the European Commission under grant agreement 951786.
CSA is supported by a studentship jointly by the UK Engineering and Physical Sciences Research Council–supported Centre for Doctoral Training in Modelling of Heterogeneous Systems, Grant No. EP/S022848/1 and the Atomic Weapons Establishment.
ABP acknowledges funding from CASTEP-USER funded by UK Research and Innovation under the grant agreement EP/W030438/1.
Calculations were performed using the Sulis Tier 2 HPC platform hosted by the Scientific Computing Research Technology Platform at the University of Warwick. Sulis is funded by EPSRC Grant EP/T022108/1 and the HPC Midlands+ consortium.
We acknowledge the University of Warwick Scientific Computing Research Technology Platform for assisting the research described within this study.


\bibliography{bib_etal.bib}
\bibliographystyle{apsrev4-1}


\end{document}